\documentclass[pra,final,titlepage,showpacs]{revtex4}
\usepackage{amsfonts}
\usepackage{graphicx}
\usepackage{amssymb}

\newcommand{\beq}{\begin{equation}}
\newcommand{\eeq}{\end{equation}}
\newcommand{\bea}{\begin{eqnarray}}
\newcommand{\eea}{\end{eqnarray}}
\newcommand{\bean}{\begin{eqnarray}}
\newcommand{\eean}{\end{eqnarray}}
\begin{document}

\title{Discrete Embedded Solitons}
\author{Kazuyuki~Yagasaki}
\affiliation{Department of Mechanical and Systems Engineering,
 Gifu University, Gifu 501-1193, Japan}
 \author{Alan R.~Champneys}
\affiliation{Department of Engineering Mathematics, University of
Bristol, Bristol BS8 1TR, UK}
\author{Boris A.~Malomed}
 \affiliation{
Department of Interdisciplinary Studies,
 School of Electrical Engineering,
 Faculty of Engineering, Tel Aviv University, Tel Aviv 69978, Israel}

\begin{abstract}
We address the existence and properties of discrete
\textit{embedded solitons} (ESs), that is localized waves existing
inside the phonon band in a nonlinear dynamical-lattice model. The
model describes a one-dimensional array of optical waveguides with
both $\chi ^{(2)}$ (second-harmonic generation) and $\chi ^{(3)}$
(Kerr) nonlinearities, for which a rich family of ESs are known to
occur in the continuum limit. First, a simple motivating problem
is considered, in which the $\chi ^{(3)}$ nonlinearity acts in a
single waveguide. An explicit solution is constructed
asymptotically in the large-wavenumber limit. The general problem
is then shown to be equivalent to the existence of a homoclinic
orbit in a four-dimensional reversible map. From properties of
such maps, it is shown that (unlike ordinary gap solitons),
discrete ESs have the same codimension as their continuum
counterparts. A specific numerical method is developed to compute
homoclinic solutions of the map, that are symmetric under a
specific reversing transformation. Existence is then studied in
the full parameter space of the problem. Numerical results agree
with the asymptotic results in the appropriate limit and suggest
that the discrete ESs may be semi-stable as in the continuous
case.
\end{abstract}


\maketitle
\section{Introduction}

An \emph{embedded soliton} (ES) is an isolated solitary wave in a
non-integrable system that resides insides the continuous spectrum
of linear waves. Unlike regular \emph{gap solitons}, the existence
of ESs in continuum models is generally of codimension-one in the
model's parameter space. That is, they are embedded into wider
families of \emph{generalized} solitary waves which have
nonvanishing periodic tails attached to them, such that the ES
occurs at isolated values of the frequency, or other intrinsic
parameter of the solution family at which the tail's amplitude
vanishes. The concept of ESs was introduced in \cite{YaMaKa:99}
(see also \cite{ChMaYaKa:01,YaMaKaCh:01} for more details),
although particular examples of waves of this kind were actually
known earlier (but not under this name)
\cite{ChGr:97,FuEs:97,ChMaFr:98}. In the last few years, ESs have
been shown to be of relevance in a diverse range of nonlinear-wave
models, see, e.g., ~\cite{KoChBuSa:02,AtMa:01,EsFuGo:03,Ya:03} and
references therein. One of their characteristic properties is that
they are typically, at best, \emph{semi-stable}. That is, the
linearization around them gives rise to no unstable eigenvalues,
but a non-trivial zero-eigenvalue mode can be found. The latter
leads to an algebraic-in-time instability for energy-decreasing
perturbations, and a similar algebraic-in-time relaxation back to
the original pulse for energy-increasing perturbations
\cite{PeYa:02}. Only in special cases of systems which have more
than one dynamical invariant, or for which the travelling-wave
system is reversible but not Hamiltonian, have examples been found
of ESs that are truly asymptotically stable \cite{Ya:03}. Also
known are physically relevant models with special symmetries that
support continuous (rather than discrete) families of ESs, and a
large part thereof may be stable \cite{Mak,Merhasin}.

All the above-cited examples pertain to continuum models. A
fundamental issue is whether solitons of the embedded type may
also occur in discrete nonlinear equations modelling dynamical
lattices. For example, in \cite{ChKi:99} it was argued that moving
topological defects in Frenkel-Kontorova (discrete sine-Gordon)
lattices can meaningfully be described as \emph{embedded kinks}.
Some of specific examples of moving discrete solitons constructed
by means of the \textit{inverse method} (a model is sought for to
which a given moving soliton is an exact solution \cite{Flach})
may also have the embedded character. Such waves, which connect a
zero state to itself, translated through a multiple of the period
of the lattice potential (the `topological charge'), were first
(indirectly) identified by Peyrard and Kruskal \cite{PeKr:84} and
have later received much attention in a wide variety of
applications \cite{KiBr:98}. In a co-moving frame, these waves
should be thought of as localized solutions to
infinite-dimensional \emph{advance and delay} (nonlocal)
equations, see works \cite{SaZoEi:00,AiChRo:03} where branches of
embedded kinks with the topological charge 1 and 2 were computed.
Another recent study \cite{ESdiscrete} reported the existence of
ESs in a discrete lattice of the Ablowitz-Ladik type, but with a
quintic nonlinearity added to the usual cubic terms, and also with
an additional next-nearest-neighbor linear coupling. By tuning the
nonlinear terms, an explicit analytical solution for a discrete ES
was found in that model (which somehow resembles the
above-mentioned inverse method). Apart from these, we are aware of
no other work that systematically considers the issue of ESs in
discrete models.

A general objective of this paper is to understand the existence
and multiplicity properties of discrete solitons in lattice
systems for which the corresponding continuum model is known to
support ESs. The first fundamental issue is whether the
codimension of their existence is preserved under discretization.
In dissipative systems, solitary pulses are supported due to the
balance between forcing and damping, hence the corresponding
homoclinic orbits are always of codimension one. Thus, formally
similar to ESs, dissipative solitons typically exist as solutions
to the corresponding stationary ordinary differential equation at
discrete values of a relevant parameter (temporal or spatial
frequency, depending on the physical realization). Then, it will
generically happen that, under the discretization of that
equation, the codimension of the solitons \emph{decreases}. This
is possible because discretization of homoclinic orbits to
hyperbolic equilibria leads to transverse homoclinic connections
that exist for a range of parameter values \cite{FiSc}. However,
for ESs we shall show that, in contrast, discrete ESs typically
keep \emph{the same} codimension as their continuum counterparts,
i.e., they also exist at discrete values of the corresponding
physical parameter.

In this work, we restrict ourselves to a discretization of a
single continuum model, for which the discrete counterpart finds a
direct physical interpretation. In fact, our results and
methodology will carry over to other discretized models that
support ESs (this expectation is supported by the general fact
that equations of the discrete nonlinear-Schr\"{o}dinger type,
that we consider below, may be derived as an asymptotic limit from
a much larger class of discrete models \cite{Morgante}).
Specifically, we shall start with the continuum model in which ESs
were identified \cite{YaMaKa:99}. It describes an optical medium
with competing quadratic ($\chi ^{(2)}$) and cubic ($\chi ^{(3)}$)
nonlinearities (see \cite{Bang1,Bang2,Bang3} and references
therein for physical applications). In this model, the evolution
variable $z$ is the propagation distance in the nonlinear optical
waveguide, and the variable $t$ is either the reduced time (in the
case of temporal solitons) or, which is more physically relevant,
a transverse spatial coordinate in a planar waveguide. In the
temporal case, such a model can be written in dimensionless form
as (see detailed explanations in reviews
\cite{Review,DiscrChi2soliton})
\begin{eqnarray}
&&i\frac{\partial u}{\partial z}+\frac{1}{2}\frac{\partial ^{2}u}{\partial
t^{2}}+u^{\ast }v+\gamma _{1}(|u|^{2}+2|v|^{2})u=0,  \nonumber \\[-1ex]
&&  \label{eqn:PDE} \\[-1ex]
&&i\frac{\partial v}{\partial z}+\frac{\delta }{2}\frac{\partial
^{2}v}{\partial t^{2}}+qv+\frac{1}{2}u^{2}+2\gamma
_{2}(|v|^{2}+2|u|^{2})v=0, \nonumber
\end{eqnarray}where $u$ and $v$ are the local amplitudes of the fundamental-frequency (FF)
and second-harmonic (SH) fields, an asterisk denotes complex
conjugation, $\delta $ is the relative dispersion of the SH, $q$
is the wavenumber mismatch, and $\gamma _{1,2}$ measure the
relative strength of the cubic (Kerr) nonlinearity compared to its
quadratic (SH-generating) counterpart, whose strength is
normalized to be $1$. In the physical model, $\gamma _{1,2} $ must
have the same sign (typically, they are positive, but may, in
principle, be negative too). The spatial variant of the system
(\ref{eqn:PDE}) takes the same form, with a partly different
interpretation of the dimensionless parameters, and is easier to
implement experimentally \cite{Bang1,Bang2,Bang3}. In the spatial
domain, one always has $\delta \equiv 1$.

The existence of ESs in this model was established in
\cite{YaMaKa:99,YaMaKaCh:01} for both cases, $\gamma _{1,2}>0$ and
$\gamma _{1,2}<0$. In the case of relevance to spatial waveguides
($\delta =1$), fundamental ESs are absent in the model, while
higher-order ones can be found.

In models such as (\ref{eqn:PDE}) which includes $\chi ^{(2)}$
interaction, ESs can be embedded only in the continuous spectrum
of the SH component, while the FF wavenumber can never be located
inside the continuous spectrum. This feature is stipulated by the
asymmetry between the two equations in system (\ref{eqn:PDE}). If
the SH supports linear waves, while the FF has the possibility of
exponential localization like $e^{-\lambda |t|}$, then the $u^{2}$
term, which drives the SH field in the second equation
(\ref{eqn:PDE}), allows it to have tails that decay exponentially
at a rate $e^{-2\lambda |t|}$. Such a solitary wave was said in
reference \cite{YaMaKa:99} to be \emph{tail-locked} and,
accordingly, the SH equation to be \emph{nonlinearisable} (because
the quadratic term can never be neglected in this equation). The
tail locking does not take place if the SH field supports its own
exponentially decaying tails that asymptotically (for
$|t|\rightarrow \infty $) dominate over the tails induced by the
term $u^{2}$.

In this paper we study the existence of discrete ESs in a model
arising from discretization of the (actually, spatial)
$t$-coordinate in (\ref{eqn:PDE}). One motivation for doing this
is to understand the effect of numerical approximation (which also
implies discretization, once a finite-difference scheme is
employed) on the computation of ESs in model systems. But this is
not our primary motivation. Lattice models play an increasingly
important role in describing physical phenomena in a number of
newly developed experimental settings. In particular, the discrete
version of the $\chi ^{(2)}:\chi ^{(3)}$ model describes an array
of the corresponding linearly-coupled waveguides. The creation of
discrete spatial solitons in a system of channel waveguides with
the $\chi ^{(2)}$ nonlinearity was recently reported
\cite{DiscrChi2soliton}. Our assumption concerning relevant
solutions is that, once the continuum version of the model readily
gives rise to ESs, then there is a chance to find them in the
corresponding lattice model too. In this paper, we show that this
is the case indeed and investigate how we may then pass to the
continuum limit. We report the corresponding discrete ES that can
be found, under special conditions, in an approximate analytical
form, and, in the general case, numerically.

The paper is organized as follows. In Section~2, we present the
lattice model to be considered in this work and discuss its
physical applicability. In Section~3, an asymptotic analysis is
undertaken for a reduced model where the cubic nonlinearity is
present only at a single site. Section~4 then goes on to discuss
the meaning of discrete ESs, as a solution to a system of
stationary finite-difference equations, in terms of a homoclinic
orbit to a fixed point of a discrete-time reversible map. This
makes it possible to understand that the solutions we seek must
have codimension one, similar to ESs in continuum models. In terms
of the same approach, in Section~5 we describe the numerical
procedure developed to search for discrete ES solutions. Note that
this requires special adaptation or other methods for finding
homoclinic orbits of maps, owing to the non-hyperbolic nature of
the fixed point. Our approach heavily relies on the reversibility
although it can be readily modified to treat non-reversible ESs.
Numerical results are reported in Section~6, in the form of the
corresponding codimension-one solution branches in the parameter
space. An array of actual shapes of the ESs is displayed too. Our
numerical results also show that the asymptotic analysis of
Section~3 can well predict the codimension-one set of ESs in the
parameter space in the model when the wavenumber is large. The
paper is concluded by Section~7 along with a preliminary stability
result for the fundamental ES.

\section{The models}

We now consider Eqs.~(\ref{eqn:PDE}) in the spatial domain. Direct
discretization of the spatial second derivative produces a lattice
model
\begin{eqnarray}
&&i\frac{du_{n}}{dz}+\frac{1}{2}D(u_{n+1}+u_{n-1}-2u_{n})+u_{n}^{\ast
}v_{n}+\gamma _{1}(|u_{n}|^{2}+2|v_{n}|^{2})u_{n}=0,  \nonumber \\[-1ex]
&&  \label{eqn:lattice} \\[-1ex]
&&i\frac{dv_{n}}{dz}+\frac{1}{2}\delta
D(v_{n+1}+v_{n-1}-2v_{n})+qv_{n}+\frac{1}{2}u_{n}^{2}+2\gamma
_{2}(|v_{n}|^{2}+2|u_{n}|^{2})v_{n}=0, \nonumber
\end{eqnarray}where $n$ is the discrete transverse coordinate which assumes integer
values, and $D=1/h^{2}$ with $h$ the stepsize of the
discretization. Effective coefficients of the lattice diffraction
for the FF and SH waves are $D_{1}\equiv D$ and $D_{2}=\delta D$;
with $\delta =1$, this yields $D_{2}=D_{1}$ in
Eqs.~(\ref{eqn:lattice}).
Unequal values of these coefficients, $D_{1}\neq D_{2}$ (i.e.,
$\delta \neq 1 $), including those with \emph{opposite signs}, can
in principle be realized experimentally (the latter possibility
was recently employed to predict the existence of discrete gap
solitons \cite{PanosZiad}). To do this one would use a
\textit{diffraction-management} technique, which is based on
oblique propagation of the beam across the waveguide array which
is modeled by the discrete system
\cite{DiffrManagement1,DiffrManagement2}. In that case, the
effective lattice-diffraction coefficients are
\begin{equation}
D_{m}=D_{m}^{(0)}\cos (mQ),\quad m=1,2,  \label{management}
\end{equation}where $Q\equiv k\sin \theta $ is the transverse wavenumber, $k$ is the
beam's wavenumber (normalization is such that the array spacing is
unity), $\theta $ is the angle between the Poynting vector and the
coordinate $z$ running along the waveguides, and $D_{1,2}^{(0)}$
are the original lattice-dispersion coefficients, corresponding to
$Q=0$. As seen in Eq.~(\ref{management}), one can efficiently
control the size and signs of $D_{1,2}$ by choosing an appropriate
angle $\theta $. In particular, the ratio $D_{2}/D_{1}$ can be
made large by choosing $Q=\pi /2+\varepsilon $, or $Q=3\pi
/2-\varepsilon $, for a small positive $\varepsilon $.

To cast the model in a normalized form, we note that $D_{1}$ can
be made positive, if it was originally negative, by means of the
complex conjugation (i.e., Eqs.~(\ref{eqn:lattice}) are replaced
by their counterparts for $u_{n}^{\ast }$ and $v_{n}^{\ast }$),
combined with the changes $v_{n}\rightarrow -v_{n}$ and $\gamma
_{1,2}\rightarrow -\gamma_{1,2}$. The size of $D_{1}$ may be set
equal to $1$ by means of the rescaling: $zD_{1}\rightarrow z$,
$\left( u_{n},v_{n}\right)/D_{1} \rightarrow \left(
u_{n},v_{n}\right)$, which leads to the final normalized form of
the discrete model,
\begin{eqnarray}
&& i\frac{du_{n}}{dz}+\frac{1}{2}(u_{n+1}+u_{n-1}-2u_{n})+u_{n}^{\ast}v_{n}
+\tilde{\gamma}_{1}(|u_{n}|^{2}+2|v_{n}|^{2})u_{n}=0,  \nonumber \\[-1ex]
&&  \label{normalised} \\[-1ex]
&& i\frac{dv_{n}}{dz}+\frac{1}{2}\delta(v_{n+1}+v_{n-1}-2v_{n})
+\tilde{q}v_{n}+\frac{1}{2}u_{n}^{2}
+2\tilde{\gamma}_{2}(|v_{n}|^{2}+2|u_{n}|^{2})v_{n}=0,  \nonumber
\end{eqnarray}
where
\begin{equation}
\tilde{q}=q/D_{1},\quad \tilde{\gamma}_{1,2}=D_{1}\gamma _{1,2}.
\label{delta}
\end{equation}
The dimensionless constants (\ref{delta}), including $\delta$, may, in
general, be positive, zero, or negative.

To obtain some analytical results, we shall also introduce a simplified
model where ESs can be found in approximate form. In the simplified model,
only the central site (the one corresponding to $n=0$) carries the Kerr
nonlinearity. Such a model is physically feasible too, if the central
waveguide in the array is doped to enhance its Kerr nonlinearity. Thus, the
simplified model is based on the equations
\begin{eqnarray}
&& i\frac{d u_{n}}{d z} +\frac{1}{2}\left(u_{n+1}+u_{n-1}-2u_{n}\right)+u_{n}^{\ast }v_{n}
+\tilde{\gamma}_{1}\epsilon_{n}\left(|u_{0}|^{2}+2|v_{0}|^{2}\right)u_{0}=0,  \nonumber \\[-1ex]
&&  \label{eqn:mod} \\[-1ex]
&& i\frac{d v_{n}}{d z} +\frac{1}{2}\delta
\left(v_{n+1}+v_{n-1}-2v_{n}\right)
+\tilde{q}v_{n}+\frac{1}{2}u_{n}^{2}
+2\tilde{\gamma}_{2}\epsilon_{n}\left(|v_{0}|^{2}+2|u_{0}|^{2}\right)v_{0}=0,
\nonumber
\end{eqnarray}
where $\epsilon _{n}=0$ for $n\neq 0$, and $\epsilon _{n}=1$ for $n=0$.

In either model, (\ref{eqn:lattice}) or (\ref{eqn:mod}), the ES
corresponds to a solution with FF component exponentially
localized:
\begin{equation}
u_{n}\sim A\exp (i\tilde{k}z-\lambda |n|)\qquad
\mbox{as
$|n|\rightarrow\infty$},  \label{uasympt}
\end{equation}where $A$ is a real amplitude and $\lambda $ is positive and real.
Simultaneously, the propagation constant $\tilde{k}$ must belong to the
\textit{phonon band} of the SH equation. That is, the SH component of the
solution generically will have a nonvanishing tail, of the form
\begin{equation}
v_{n}\sim C\exp (2i\tilde{k}z-ip|n|)\qquad \mbox{as $|n|\rightarrow \infty$},
\label{vasympt}
\end{equation}where $p$ is real. The existence of an ES corresponds to the vanishing of
the constant $C$ in Eq.~(\ref{vasympt}). Linearization of
Eqs.~(\ref{eqn:lattice}) and (\ref{eqn:mod}) and subsequent
substitution of expressions (\ref{uasympt}) and (\ref{vasympt})
yield the following relations:
\begin{equation}
\tilde{k}=2\sinh ^{2}(\lambda /2),  \label{k}
\end{equation}\begin{equation}
\tilde{q}-2\tilde{k}=2\delta \sin ^{2}(p/2).  \label{q}
\end{equation}From Eqs.~(\ref{k}) and (\ref{q}) we see that a discrete ES may exist in the
case of $\tilde{k}>0$, with $\tilde{q}$ belonging to the region
\begin{equation}
0<\tilde{q}-2\tilde{k}<2\delta \quad \mathrm{or}\quad 2\delta
<\tilde{q}-2\tilde{k}<0,  \label{area}
\end{equation}depending on whether $\delta $ is positive or not. Note that we can only
choose $\tilde{k}$, $\tilde{q}$ and $\delta $ independently in the
set of $(\tilde{k},\tilde{q},\delta ,\lambda ,p)$ since $\lambda $
and $p$ are determined by the other three parameters via Eqs.
(\ref{k}) and (\ref{q}).

\section{Asymptotic analysis for the simplified model}

We first consider the simplified model (\ref{eqn:mod}), with the
$\chi ^{(3)} $ nonlinearity present only at the central site. An
ES solution is sought asymptotically under the assumption that
$|v_{n}|\ll |u_{n}|=O(1)$, the validity of which will be checked
\textit{a posteriori}. Accordingly, the quadratic term in the
first equation of the system (\ref{eqn:mod}) may be dropped to
first approximation, which makes the expressions (\ref{uasympt})
and (\ref{k}) an \emph{exact solution} for $n\neq 0$. At the point
$n=0$, the cross-phase-modulation term, $|v_{0}|^{2}u_{0}$, in the
first equation (\ref{eqn:mod}) may be dropped too to leading
order. Then, the equation at $n=0$ yields a final result for the
FF component of the soliton, $A^{2}=\tilde{\gamma}_{1}^{-1}\sinh
\lambda $, which implies that the discrete soliton in the FF
component is supported by itself, without coupling to the SH
component. Further, using Eq.~(\ref{k}) one can express $A^{2}$ in
terms of $\tilde{k}$,
\begin{equation}
A^{2}=\tilde{\gamma}_{1}^{-1}\sqrt{\tilde{k}(\tilde{k}+2)},  \label{Ak}
\end{equation}
which means that the solution exists only in the case of
$\tilde{\gamma}_{1}>0$.

Now, we tackle the second equation of (\ref{eqn:mod}) and
substitute the FF field in the form of Eqs.~(\ref{uasympt}) and
(\ref{k}). Obviously, at $n\neq 0$, an exact solution can be
found, which precisely corresponds to an ES, as it does not
contain the nonvanishing tail (\ref{vasympt}) and is `tail-locked'
to the square of the FF field, as explained in the Introduction.
We find explicitly that
\begin{equation}
v_{n}=B\exp (2i\tilde{k}z-2\lambda |n|),  \label{vsolution}
\end{equation}where
\begin{equation}
B=-\frac{A^{2}}{2[2\delta \sinh ^{2}\lambda
+(\tilde{q}-2\tilde{k})]}\equiv
-\frac{1}{2\tilde{\gamma}_{1}}\frac{\sqrt{\tilde{k}(\tilde{k}+2)}}{2\delta
\cdot \tilde{k}(\tilde{k}+2)+(\tilde{q}-2\tilde{k})}.  \label{B}
\end{equation}Substituting the expressions (\ref{uasympt}) and (\ref{vsolution}) into the
second equation of (\ref{eqn:mod}) at $n=0$, we have
\begin{equation}
\cosh \lambda =\frac{2\tilde{\gamma}_{2}}{\delta \tilde{\gamma}_{1}}.
\label{eqn:lambda}
\end{equation}Here, following the assumption $|v_{n}|\ll |u_{n}|$, we neglected the
self-phase-modulation term $|v_{0}|^{2}v_{0}$ at this point.
Equation~(\ref{eqn:lambda}) also implies that
\begin{equation}
\frac{2\tilde{\gamma}_{2}}{\delta \tilde{\gamma}_{1}}>1,  \label{eqn:12}
\end{equation}and, especially, that $\delta $ has the same sign as $\tilde{\gamma}_{2}$
since $\lambda ,\,\tilde{\gamma}_{1}>0$. Finally, using the
relations (\ref{k}) and (\ref{eqn:lambda}), we obtain
\begin{equation}
\tilde{k}=\frac{2}{\delta }\frac{\tilde{\gamma}_{1}}{\tilde{\gamma}_{2}}-1,
\label{final}
\end{equation}which is positive by virtue of (\ref{eqn:12}).
This result selects the \emph{single value} of $\tilde{k}$ at which the simplified model admits the
existence of an ES, with tail-locked SH component. It follows from
Eq.~(\ref{final}) that such a solution exists if the relative
lattice-diffraction coefficient (see Eq.~(\ref{delta})) belongs to
the interval
\begin{equation}
0<|\delta |<2\tilde{\gamma}_{1}/|\tilde{\gamma}_{2}|,\quad
\tilde{\gamma}_{1}>0,  \label{12+}
\end{equation}where $\delta $ and $\tilde{\gamma}_{2}$ must be of the same sign.

It is now necessary to check compatibility of the solution with
the underlying assumption, $|v_{n}|\ll |u_{n}|$, which implies
$A^{2}\ll B^{2}$ (see Eqs.~(\ref{Ak}) and (\ref{B})).
Straightforward consideration shows that this condition amounts to
$|\delta|\ll|\tilde{\gamma}_{1,2}|$. We also note that, if the FF
component of the soliton is strongly localized, so that
$e^{-\lambda }\ll 1$ (see Eq.~(\ref{uasympt})), the simplified
model is actually equivalent to the full one, as the nonlinearity
in the first equation of (\ref{eqn:lattice}) is then negligible at
$n\neq 0$. From (\ref{k}) we see that this condition implies that
$\tilde{k}$ must be large and hence, from (\ref{final}), that
$|\delta|\ll 1$.

A noteworthy feature of Eqs.~(\ref{final}) and (\ref{12+}) is that
they do not involve the mismatch parameter $\tilde{q}$. However,
the condition that the soliton found above is embedded implies
that $\tilde{k}$ given by Eq.~(\ref{final}) must belong to the
interval (\ref{area}), which relates $\tilde{k}$ to $\tilde{q}$,
as $|\delta|\ll 1$. This approximate analysis, valid too for the
original model in the limit of small $|\delta|$, means that, for
an isolated selected $\tilde{k}$-value given by (\ref{final}),
there exists a curve of single-humped ESs approximately
parametrized by $\tilde{q}$ in the narrow interval
\begin{equation}
\tilde{q}\in (2\tilde{k},2\tilde{k}+2\delta) \quad\mbox{for
$\delta>0$}\quad\mbox{or}\quad (2\tilde{k}+2\delta,2\tilde{k})
\quad\mbox{for $\delta<0$} \label{qinterval}
\end{equation}
in the $(\tilde{q},\tilde{k})$-plane for $\delta$ fixed.

\section{Reversible maps}

Let us now consider the model (\ref{eqn:lattice}) in the general
case, and understand, from a dynamical systems point of view, why
discrete ESs should exist. In this context, one of essential
issues is the transition to the continuum limit. The scaling
leading to Eq. (\ref{normalised}), in which $D_{1}=1$, does not
allow this. Hence, we undo this scaling and set $D_{1}=D$,
$D_{2}=\delta D$. As above, we look for stationary solutions in
the form
\begin{equation}
u_{n}=U_{n}e^{ikz},\quad v_{n}=V_{n}e^{2ikz},  \label{eqn:UV}
\end{equation}where $U_{n}$ and $V_{n}$ are real. Equations~(\ref{eqn:lattice}) thus
reduce to
\begin{eqnarray}
&&\frac{1}{2}D(U_{n+1}+U_{n-1}-2U_{n})-kU_{n}+U_{n}V_{n}+\gamma
_{1}(U_{n}^{2}+2V_{n}^{2})U_{n}=0,  \nonumber \\[-1ex]
&&  \label{eqn:disc} \\[-1ex]
&&\frac{1}{2}\delta
D(V_{n+1}+V_{n-1}-2V_{n})+(q-2k)V_{n}+\frac{1}{2}U_{n}^{2}+2\gamma
_{2}(V_{n}^{2}+2U_{n}^{2})V_{n}=0,  \nonumber
\end{eqnarray}where parameters without the tildes are related to those with tildes by
\begin{equation}
q=\tilde{q}D,\quad k=\tilde{k}D,\quad \gamma _{1,2}=\tilde{\gamma}_{1,2}/D
\label{rescaledpars}
\end{equation}Note that the possible existence regions of ESs, given by expression (\ref{area}), now becomes
\begin{equation}
0<q-2k<2\delta D\quad \mathrm{or}\quad 2\delta D<q-2k<0.  \label{area_scale}
\end{equation}

First and foremost, we want to establish the codimension of ES
solutions to (\ref{eqn:disc}). In order to do that, it is useful
to recast the system as a four-dimensional map. Specifically, upon
scaling the variables as $\xi _{n}\equiv \sqrt{2|\gamma _{1}|/D}\
U_{n}$ and $\eta _{n}\equiv \sqrt{2|\gamma _{1}|/D}\ V_{n}$, we
can indeed view Eqs.~(\ref{eqn:disc}) as defining a\emph{\
four-dimensional map},
\begin{equation}
\zeta _{n+1}=F_{\epsilon }(\zeta _{n}),  \label{eqn:F}
\end{equation}
where $\zeta _{n}\equiv (\chi _{n},\xi _{n},\mu _{n},\eta _{n})$ and
\[
F_{\epsilon }(\zeta _{n})=\left(
\begin{array}{c}
\xi _{n} \\
f_{\epsilon }^{(1)}(\zeta _{n}) \\
\eta _{n} \\
f_{\epsilon }^{(2)}(\zeta _{n})\end{array}\right) .
\]
Here we define
\begin{eqnarray}
f_{\epsilon }^{(1)}(\zeta _{n}) &=&-\chi _{n}+2\nu _{1}\xi _{n}-(\epsilon
\xi _{n}\eta _{n}+\kappa _{1}(\xi _{n}^{2}+2\eta _{n}^{2})\xi _{n}), \\
f_{\epsilon }^{(2)}(\zeta _{n}) &=&-\mu _{n}+2\nu _{2}\eta _{n}-\delta
^{-1}\left( \frac{1}{2}\epsilon \xi _{n}^{2}+2\kappa _{2}(\eta _{n}^{2}+2\xi
_{n}^{2})\eta _{n}\right) ,
\end{eqnarray}
where
\begin{eqnarray}
\nu _{1} &=&1+\frac{k}{D},\quad \nu _{2}\equiv 1+\frac{2k-q}{\delta D},\quad
\epsilon \equiv \sqrt{\frac{2}{D|\gamma _{1}|}},  \nonumber \\[-1.5ex]
&&  \label{eqn:para} \\[-1.5ex]
\kappa _{1} &\equiv &{\mathrm{sgn}}(\gamma _{1}),\quad \kappa _{2}\equiv
\gamma _{2}/\left\vert \gamma _{1}\right\vert .  \nonumber
\end{eqnarray}

The map $F_{\epsilon }$ is \textit{reversible} \cite{De:76b,LaRo:98}, in the
sense that $F_{\epsilon }^{-1}(R(\zeta _{n}))=R(F_{\epsilon }(\zeta _{n}))$
holds, where $R:\mathbb{R}^{4}\rightarrow \mathbb{R}^{4}$ is the (linear)
involution given by
\begin{equation}
R:(\chi _{n},\xi _{n},\mu _{n},\eta _{n})\mapsto (\xi _{n},\chi _{n},\eta
_{n},\mu _{n}).  \label{R1}
\end{equation}
Note that the map $F_{\epsilon }$ is also reversible, under the second
reversibility
\begin{equation}
\hat{R}:(\chi _{n},\xi _{n},\mu _{n},\eta _{n})\mapsto (f_{\epsilon
}^{(1)}(\zeta _{n}),\xi _{n},f_{\epsilon }^{(2)}(\zeta _{n}),\eta _{n}).
\label{R2}
\end{equation}
In what follows, we consider orbits of the map that are reversible
with respect to $R$. These will correspond to ESs that have their
central peaks on a lattice site, which are the kind of lattice
solitons that are most frequently observed in physical
applications. In contrast, ESs that are symmetric under $\hat{R}$
would have their central peak between two lattice sites. Both $R$
and $\hat{R}$ arise from the fact that the ordinary differential
equations for steady solutions of the continuum model
(\ref{eqn:PDE}),
\begin{eqnarray}
&&\frac{1}{2}\frac{d^{2}U}{dt^{2}}-kU+UV+\gamma _{1}(U^{2}+2V^{2})U=0,
\nonumber \\[-1ex]
&&  \label{eqn:cont} \\[-1ex]
&&\frac{\delta }{2}\frac{d^{2}V}{dt^{2}}+(q-2k)V+\frac{1}{2}U^{2}+2\gamma
_{2}(V^{2}+2U^{2})V=0,  \nonumber
\end{eqnarray}
are reversible too, under an involution
$\tilde{R}:(U,dU/dt,V,dV/dt)\mapsto(U,-dU/dt,V,-dV/dt)$.

A fundamental characteristic of reversible maps is that if
$\{\zeta_{n}\}_{n=-\infty }^{\infty }$ is an orbit then
$\{R(\zeta_{-n})\}_{n=-\infty }^{\infty }$ is also an orbit. We
say that an orbit $\{\zeta _{n}\}_{n=-\infty}^{\infty}$ is
\textit{symmetric} (with respect to the reversibility) if
$\zeta_{n+1}=R(\zeta _{-n})$. Denote $\mathrm{Fix}(R)=\{\zeta \in
\mathbb{R}^{4}\,|\,F_{\epsilon }(\zeta )=R(\zeta )\}$. We easily
see that $\zeta =(\chi ,\xi ,\mu ,\eta )\in \mathrm{Fix}(R)$ if
and only if $\chi =f_{\epsilon }^{(1)}(\zeta )$ and $\mu
=f_{\epsilon }^{(2)}(\zeta )$. Thus, $\mathrm{Fix}(R)$ depends on
the particular form of the map $F_{\epsilon }$. An orbit $\{\zeta
_{n}\}_{n=-\infty }^{\infty }$ is symmetric if and only if $\zeta
_{0}\in \mathrm{Fix}(R)$, i.e., $\xi _{-1}=\xi _{1}$ and $\eta
_{-1}=\eta _{1}$.

The map $F_{\epsilon }$ has a fixed point at the origin $O$, whose Jacobian
matrix is
\[
J=\left(
\begin{array}{cccc}
0 & 1 & 0 & 0 \\
-1 & 2\nu _{1} & 0 & 0 \\
0 & 0 & 0 & 1 \\
0 & 0 & -1 & 2\nu _{2}\end{array}\right) .
\]It follows from Eq.~(\ref{area_scale}) that $\nu _{1}>1$ and $|\nu _{2}|<1$,
hence the origin is a fixed point of $F_{\epsilon }$ of
saddle-center type. The saddle-center has one-dimensional stable
and unstable manifolds, $W^{s}(O)$ and $W^{u}(O)$, which are
tangent to the stable and unstable subspaces spanned by the
vectors $(1,\nu _{1}-\sqrt{\nu _{1}^{2}-1},0,0)$ and $(1,\nu
_{1}+\sqrt{\nu _{1}^{2}-1},0,0)$, respectively, and a
two-dimensional center manifold, $W^{c}(O)$, which is tangent to
the center subspace spanned by a set of two vectors, $(0,0,1,0)$
and $(0,0,0,1)$. By the fundamental property of reversible maps,
$W^{s}(O)=R(W^{u}(O))$ and $W^{u}(O)=R(W^{s}(O))$. Thus, if there
exists an orbit $\{\zeta _{n}\}_{n=-\infty }^{\infty }$ on
$W^{u}(O)$ such that $\zeta _{0}\in \mathrm{Fix}(R)$, then it is
also contained in $W^{s}(O)$ and is a symmetric homoclinic orbit
to $O$.

If $F_{\epsilon }$ were not reversible, then such intersections
between the one-dimensional stable and unstable manifolds in the
four-dimensional phase space would be of codimension two. However,
symmetric homoclinic orbits are of codimension one, since, for
their existence, we require an intersection between the
one-dimensional unstable manifold $W^{u}(O)$ and the
two-dimensional manifold $\mathrm{Fix}(R)$. Thus, since a
homoclinic orbit to $O$ for $F_{\epsilon }$ represents precisely
an ES in the lattice system (\ref{eqn:disc}), we have established
that:

\noindent \emph{Embedded solitons of the lattice model are of
\textbf{codimension-one} in the parameter space, provided they are
symmetric under a reversibility equivalent to $R$ (or $\hat{R}$)}.

\noindent Note that this is identical to the multiplicity result known in
the continuum version of the model \cite{ChMaYaKa:01}.

Finally, in what follows we shall also treat the case of pure
$\chi ^{(2)}$ nonlinearity, $\gamma _{1}=\gamma _{2}=0$. This case
is physically important, as experiments could be quite conceivably
be set up in a medium with negligible Kerr nonlinearity. However,
without the $\chi^{(3)}$ terms, no ES exists in the continuum
model \cite{ChMaYaKa:01}. It will therefore be important to find
out whether the same is true in the discrete model too.

With $\gamma _{1}=\gamma _{2}=0$, the scaling leading to $F_{\epsilon}$
becomes invalid, so we shall consider instead a family of four-dimensional
maps
\begin{equation}
\zeta _{n+1}=G_{\epsilon }(\zeta _{n}),  \label{eqn:G}
\end{equation}
where we define
\[
G_{\epsilon }(\zeta _{n})=\left(
\begin{array}{c}
\xi _{n} \\
g_{\epsilon }^{(1)}(\zeta _{n}) \\
\eta _{n}, \\
g_{\epsilon }^{(2)}(\zeta _{n})\end{array}\right) ,
\]
with
\begin{eqnarray}
g_{\epsilon }^{(1)}(\zeta _{n}) &=&-\chi _{n}+2\nu _{1}\xi _{n}-(\epsilon
\xi _{n}\eta _{n}+(1-\epsilon )(\xi _{n}^{2}+2\eta _{n}^{2})\xi _{n}), \\
g_{\epsilon }^{(2)}(\zeta _{n}) &=&-\mu _{n}+2\nu _{2}\eta _{n}-\delta
^{-1}\left( \frac{1}{2}\epsilon \xi _{n}^{2}+2(1-\epsilon )(\eta
_{n}^{2}+2\xi _{n}^{2})\eta _{n}\right),
\end{eqnarray}
with $\nu _{1}$ and $\nu _{2}$ given by (\ref{eqn:para}). The map
$G_{\epsilon }$ is reversible under the same involutions $R$ and
$\hat{R}$. The map $G_{1}$ is equivalent to (\ref{eqn:disc}) with
$\gamma _{1}=\gamma _{2}=0$ if one sets $\xi _{n}=U_{n}$ and $\eta
_{n}=V_{n}$, and simultaneously $G_{0}$ coincides with $F_{0}$
with $\kappa_{1}=\kappa _{2}=1 $. The origin $O$ is also a
saddle-center of $G_{\epsilon }$ and has the same stable, unstable
and center subspaces as those of $F_{\epsilon }$. So we can apply
all the arguments presented above for $F_{\epsilon }$ to
$G_{\epsilon }$ if we replace $F_{\epsilon }$ with $G_{\epsilon }$
in the definition of $\mathrm{Fix}(R)$. In particular, a
homoclinic orbit to $O$ for $G_{\epsilon }$ represents an ES in
the lattice system (\ref{eqn:lattice}) with $\gamma _{1}=\gamma
_{2}=0$.

\section{Numerical procedure}

Several numerical procedures exist for finding homoclinic orbits to fixed
points in maps, see, e.g., \cite{K81,BK97a,BK97b,Y98a,BBV00}. Here we shall
use an adaptation of these methods that takes into regard both the
reversibility and the non-hyperbolic nature of the fixed point.

To compute symmetric homoclinic orbits for $F_{\epsilon }$, we consider the
three-dimensional algebraic problem
\begin{eqnarray}
&&\chi _{-N}-\left( \nu _{1}-\sqrt{\nu _{1}^{2}-1}\right) \xi _{-N}=0,
\label{eqn:con1} \\
&&\xi _{1}=\xi _{-1},\quad \eta _{1}=\eta _{-1},  \label{eqn:con2}
\end{eqnarray}
for $N>0$ sufficiently large, where $\zeta_{\pm 1}
\equiv(\chi_{\pm 1},\xi_{\pm 1},\mu_{\pm 1},\eta_{\pm 1})
=F_{\epsilon }^{N\pm 1}(\chi_{-N},\xi_{-N},0,0)$. The condition
(\ref{eqn:con1}) means that the point $(\chi _{-N},\xi _{-N},0,0)$
lies in the one-dimensional unstable subspace of the fixed point
at the origin, and condition (\ref{eqn:con2}) means that $\zeta
_{0}=(\chi _{0},\xi _{0},\mu_{0},\eta _{0})\in \mathrm{Fix}(R)$,
i.e., the finite orbit $\{\zeta_{n}\}_{n=-N}^{0}$ intersects
$\mathrm{Fix}(R)$ at $n=0$. Thus, adding a parameter as an extra
unknown variable, Eqs.~(\ref{eqn:con1}) and (\ref{eqn:con2})
represent a formally well-posed system of three equations for two
unknowns, $\chi _{-N}$ and $\xi _{-N}$, and the free parameter,
which can be chosen to be any of $\epsilon$, $\delta$, $k$, $D$,
$q$, $\kappa_{1}$ or $\kappa_{2}$. A non-trivial solution for $N$
sufficiently large gives an approximate homoclinic orbit $\{\zeta
_{n}\}_{n=-N}^{N+1}$ of $F_{\epsilon }$, that is symmetric under
$R$, which in turn corresponds to a discrete ES solution of the
lattice equation (\ref{eqn:lattice}) for $\gamma _{1,2}\neq 0$. In
the case $\gamma _{1,2}=0$, the same treatment can be used to find
symmetric homoclinic orbits for $G_{\epsilon }$. Note that one
test of the validity of this approximation is to measure the
distance of the point $(\chi _{-N},\xi _{-N},0,0)$ from the
origin. By virtue of the stable manifold theorem for maps
\cite{GH83}, we know that the error will be proportional to this
distance squared.

\begin{figure}[t]
\begin{center}
\includegraphics[scale=1.2]{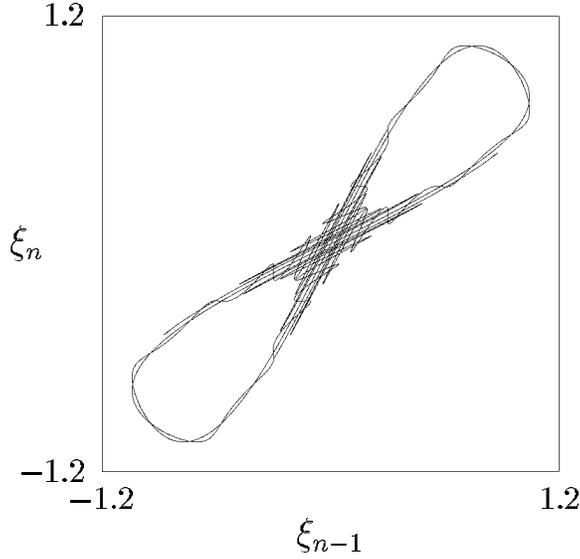}
\end{center}
\caption{Homoclinic tangles for the two-dimensional map $f$ with
$\protect\nu_1=1.25$ and $\protect\kappa _{1}=1$, drawn using the
software \texttt{Dynamics} \protect\cite{NY97}. The circle
``$\bullet$'' represents the saddle at the origin.} \label{fig:3a}
\end{figure}

Branches of solutions to Eqs.~(\ref{eqn:con1}) and
(\ref{eqn:con2}) can be continued in a second parameter using
pseudo-arclength continuation. To this end, we use the code
\texttt{AUTO} \cite{Auto}. The problem now is finding a good
initial point along a branch of discrete ESs. Here we can use a
regular perturbation approach by first finding solutions to the
map $F_{0}$, making use of the fact that when $\epsilon =0$, the
$(\chi _{n},\xi _{n})$-plane is invariant under $F_{\epsilon }$.
The restriction of $F_{0}$ onto the invariant plane is given by
\begin{equation}
(\chi _{n+1},\xi _{n+1})=f(\chi _{n},\xi _{n}),  \label{eqn:2d}
\end{equation}
where
\[
f(\chi _{n},\xi _{n})\equiv (\xi _{n},-\chi _{n}+2\nu _{1}\xi _{n}-\kappa
_{1}\xi _{n}^{3}).
\]The two-dimensional map $f$ also has a hyperbolic saddle at the origin, and
is reversible under an involution,
\[
\bar{R}:(\chi _{n},\xi _{n})\mapsto (\xi _{n},\chi _{n}).
\]The stable (resp. unstable) manifold of the saddle, $\bar{W}^{s}(O)$
(resp. $\bar{W}^{u}(O)$), is tangent to its stable (resp. unstable) subspace spanned
by a vector $(1,\nu _{1}-\sqrt{\nu _{1}^{2}-1})$ (resp.
$(1,\nu_{1}+\sqrt{\nu _{1}^{2}-1}$)). By the reversibility,
$\bar{W}^{s}(O)=\bar{R}(\bar{W}^{u}(O))$ and vice versa. When $\nu
_{1}>0$ and $\kappa _{1}>0$, the stable and unstable manifolds
intersect transversely and form homoclinic tangles, as shown in
Fig.~\ref{fig:3a}.

Let $N>0$ be a sufficiently large integer, and let
$(\bar{\chi}_{-N},\bar{\xi}_{-N})$ be a point on the unstable
subspace such that
$(\bar{\xi}_{N+1},\bar{\chi}_{N+1})=(\bar{\chi}_{-N},\bar{\xi}_{-N})$,
where
$(\bar{\chi}_{N+1},\bar{\xi}_{N+1})=f^{2N+1}(\bar{\chi}_{-N},\bar{\xi}_{-N})$.
By the reversibility of $f$, the point
$(\bar{\chi}_{N+1},\bar{\xi}_{N+1})$ must be contained in the
stable subspace. The two points $(\bar{\chi}_{-N},\bar{\xi}_{-N})$
and $(\bar{\chi}_{N+1},\bar{\xi}_{N+1})$ are also close to the
saddle when $N>0$ is large. Hence, the orbit leaving at
$(\bar{\chi}_{-N},\bar{\xi}_{-N})$ on the unstable subspace and
arriving at $(\bar{\chi}_{N+1},\bar{\xi}_{N+1})$ on the stable
subspace is a good approximation to a symmetric homoclinic orbit
of $f$. Using an adaptation of the driver \texttt{HomMap}
\cite{Y98a,Y98b} to \texttt{AUTO97}, we can easily find such
approximate homoclinic points.

\begin{figure}[tbp]
\begin{center}
\includegraphics[scale=0.7]{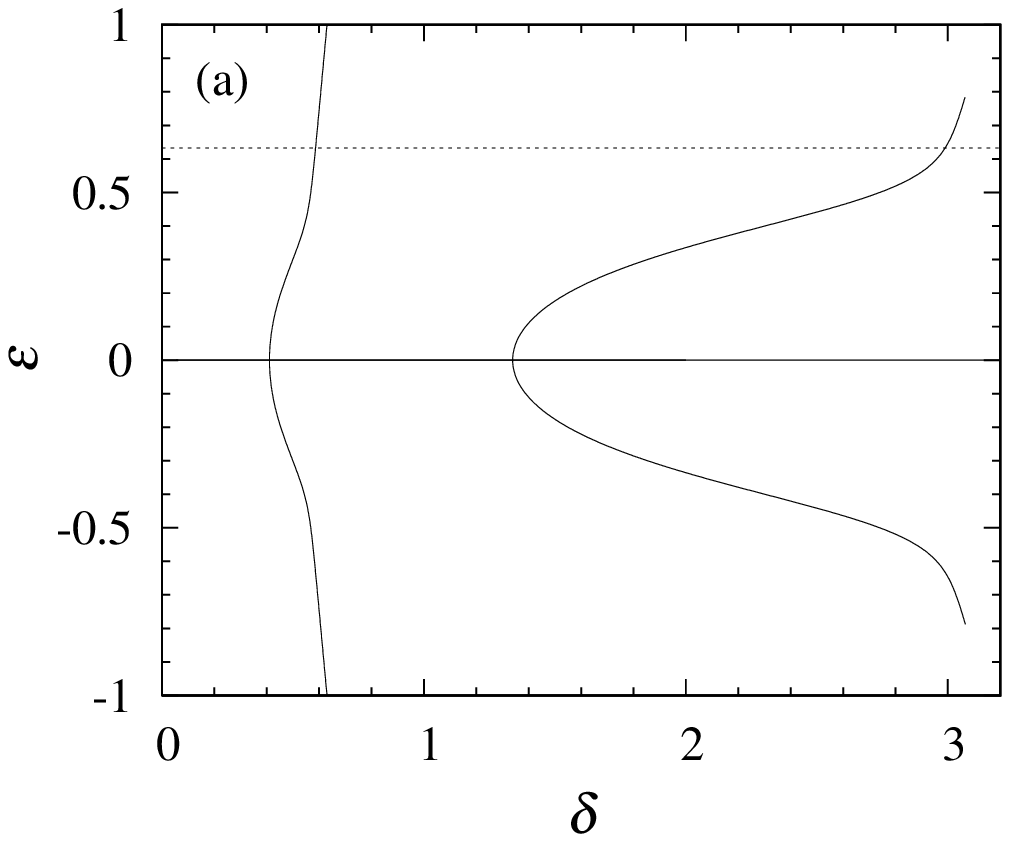}\quad \includegraphics[scale=0.7]{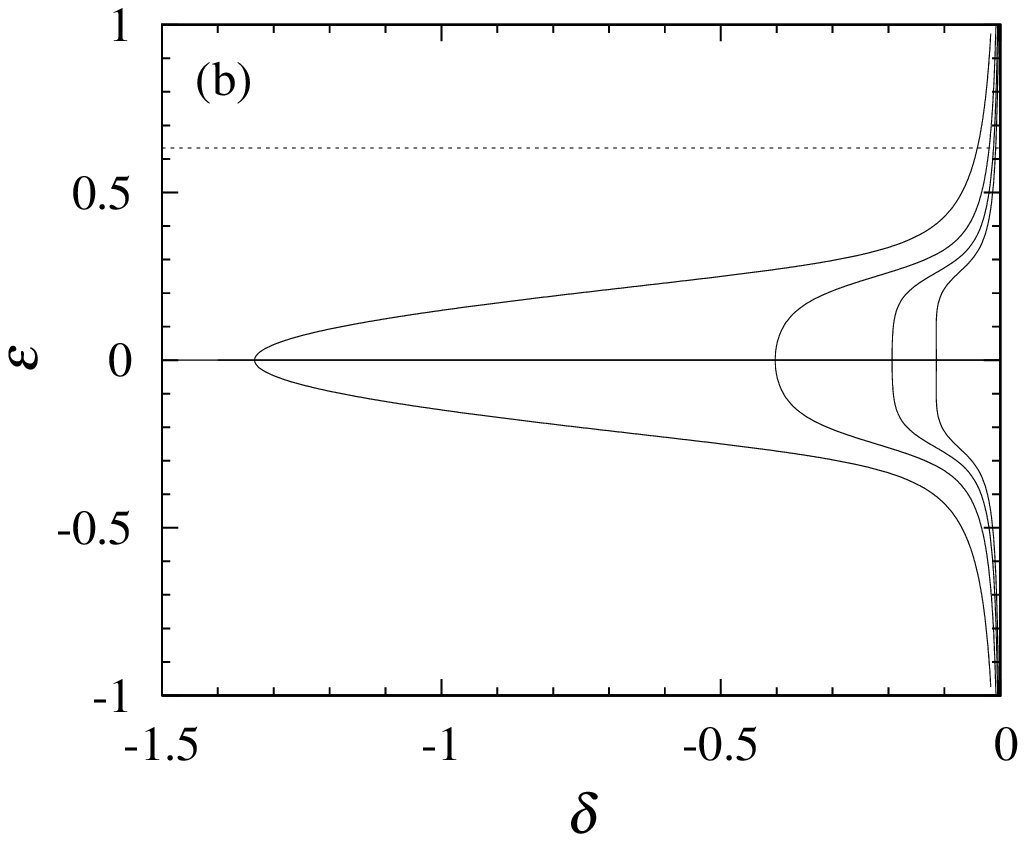}
\end{center}
\caption{Numerical continuation of symmetric homoclinic orbits for
$F_{\protect\epsilon }$ when $D=100$ and $N=85$: (a) $k=0.3$,
$q=5$ and $\protect\kappa _{1}=\protect\kappa _{2}=1$; (b)
$k=0.5$, $q=1$, $\protect\kappa _{1}=1$ and $\protect\kappa
_{2}=-1$. Here $\protect\epsilon $ and $\protect\delta $ are
varied. The dotted line represents $\protect\epsilon
=\protect\sqrt{2/(D|\protect\gamma _{1}|)}$ with $\protect\gamma
_{1}=0.05$. } \label{fig:3b}
\end{figure}

\begin{figure}[tbp]
\begin{center}
\includegraphics[scale=0.7]{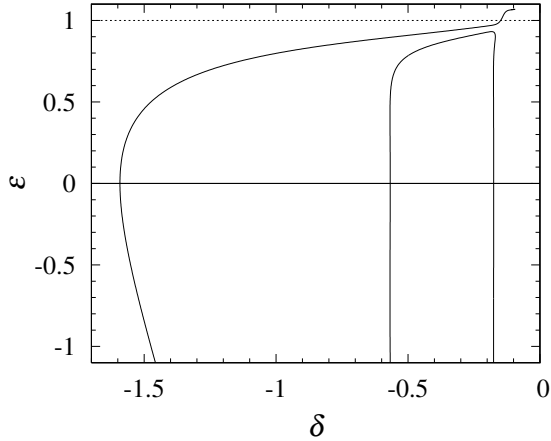}
\end{center}
\caption{Numerical continuation of symmetric homoclinic orbits for
$G_{\protect\epsilon }$ when $D=10$, $k=3$, $q=1$ and $N=15$. Here
$\protect\epsilon $ and $\protect\delta $ are varied. The dotted
line represents $\protect\epsilon =1$. } \label{fig:3c}
\end{figure}

\begin{figure}[tbp]
\begin{center}
\includegraphics[scale=0.7]{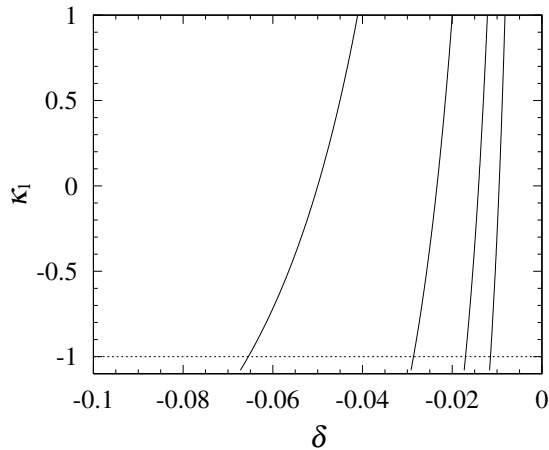}
\end{center}
\caption{Numerical continuation of symmetric homoclinic orbits for
$F_{\protect\epsilon }$ when $D=100$, $k=0.5$, $q=1$,
$\protect\kappa _{2}=-1 $, $N=85$ and $\protect\epsilon
=0.6324555\approx \protect\sqrt{2/\left( D|\protect\gamma
_{1}|\right) }$ with $\protect\gamma _{1}=0.05$. Here
$\protect\delta $ and $\protect\kappa _{1}$ are varied. The dotted
line represents $\protect\kappa_{1}=-1$. } \label{fig:3d}
\end{figure}

Now, in order to find non-trivial symmetric homoclinic solutions
of $F_{\epsilon }$ for $\epsilon >0$, we take $\epsilon $ as the
additional free parameter and choose $(\chi _{-N},\xi
_{-N},\epsilon )=(\bar{\chi}_{-N},\bar{\xi}_{-N},0)$ as the
starting solution to (\ref{eqn:con1}) and (\ref{eqn:con2}), where
$(\bar{\chi}_{-N},\bar{\xi}_{-N})$ denotes the homoclinic point on
the unstable subspace for $f$, obtained using the above procedure.
Fixing $\kappa _{1}=1$, we then performed continuation of these
algebraic equations in \texttt{AUTO}, using $\delta$ as the
continuation parameter. To obtain symmetric homoclinic orbits for
$\kappa _{1}=-1$, we also take $\delta $ and $\kappa _{1}$ as the
free and continuation parameters, respectively, and continue the
solution obtained above for $\kappa _{1}=1$ and $\epsilon
=\sqrt{2/(D|\gamma _{1}|)}$. The results are presented in
Figs.~\ref{fig:3b}-\ref{fig:3d}. As shown in Figs.~\ref{fig:3b}
and \ref{fig:3c}, new branches bifurcate from the one with
$\epsilon=0$ at discrete values of $\delta$ and can be continued
to $\epsilon=\sqrt{2/(D|\gamma_{1}|)} $ by varying $\delta $ and
$\epsilon $. As shown in Fig.~\ref{fig:3d}, the branches were also
continued from $\kappa _{1}=1$ to $\kappa _{1}=-1$ by varying
$\delta $ and $\kappa_{1}$. For $\kappa _{1}=-1$ and $\kappa
_{2}=1$, we could not find such symmetric homoclinic orbits of
$F_{\epsilon }$ with sufficient precision. In these computations,
we also chose the value of $N$ such that the distance between the
approximate homoclinic point and the origin was $1.5\times
10^{-3}$ at most, and checked that the results did not change
significantly under increase of $N$. Our computations also
suggested that a possibly infinite number of branches of symmetric
homoclinic orbits could be obtained when $k$ or $q$ is large or
$\delta $ is small. These branches show oscillations in the
parameter plane, which are sensitive to the value of $N$ and do
not appear to converge as $N\rightarrow \infty$. These, probably
spurious branches, result from a large ratio between the imaginary
eigenvalue of the linearization and the real eigenvalue, which is
well known to be a singular limit for equations that bear ES
\cite{Ch:01,Lo:01}, and needs to be treated with great caution. In
the results that follow, we have stopped computation at points
where such oscillations first become evident, and have checked all
results for consistency in the limit of $N\rightarrow \infty $.

\section{Numerical results}

We now present continuation results obtained with \texttt{AUTO}
for symmetric homoclinic orbits of $F_{\epsilon }$ (or
$G_{\epsilon }$) under simultaneous variation of two relevant
parameters within the saddle-center parameter regions. By varying
the discreteness coefficient $D$ up to large values, we are also
able to compare the results to those of the continuum limit,
$D\rightarrow \infty $. We shall treat the cases $\delta >0$ and
$\delta <0$ separately and also discuss the possibility of
discrete ESs in the absence of cubic nonlinearity.

\subsection{The case of\/ $\protect\delta >0$}

\begin{figure}[tbp]
\begin{center}
\includegraphics[scale=0.7]{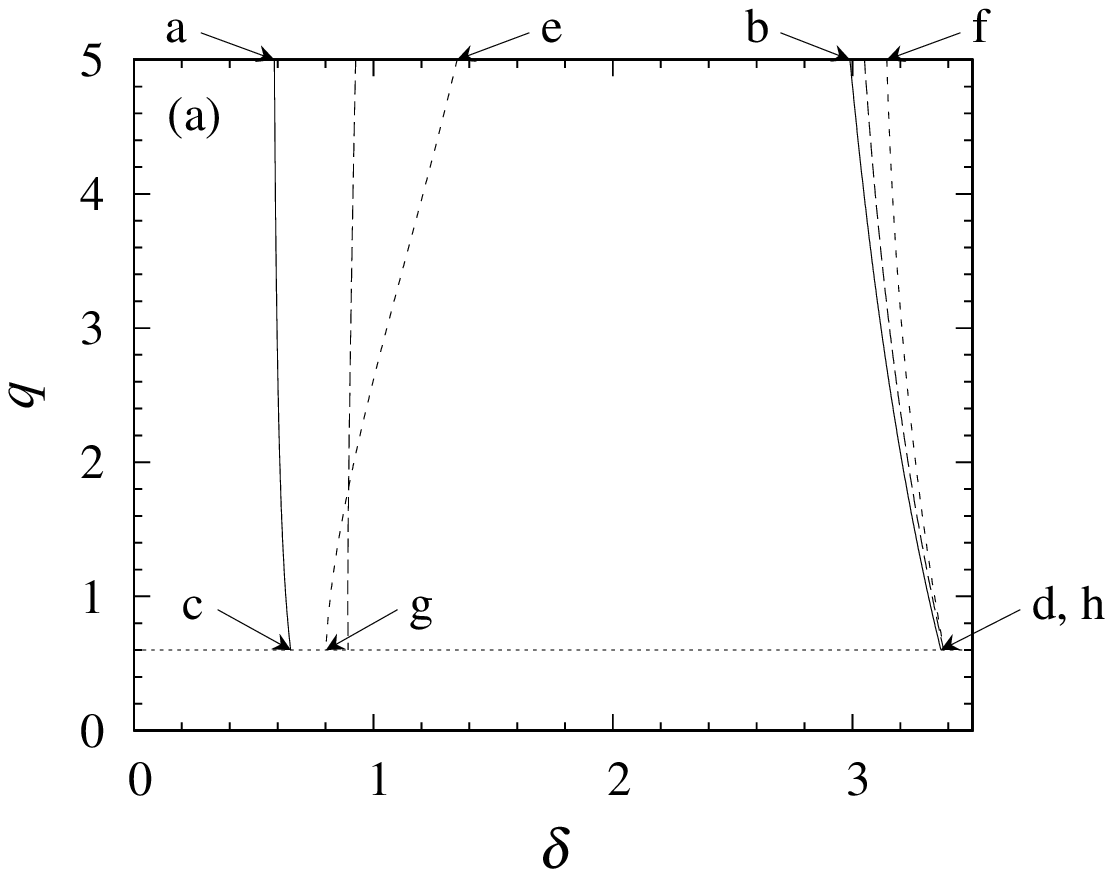}\quad \includegraphics[scale=0.7]{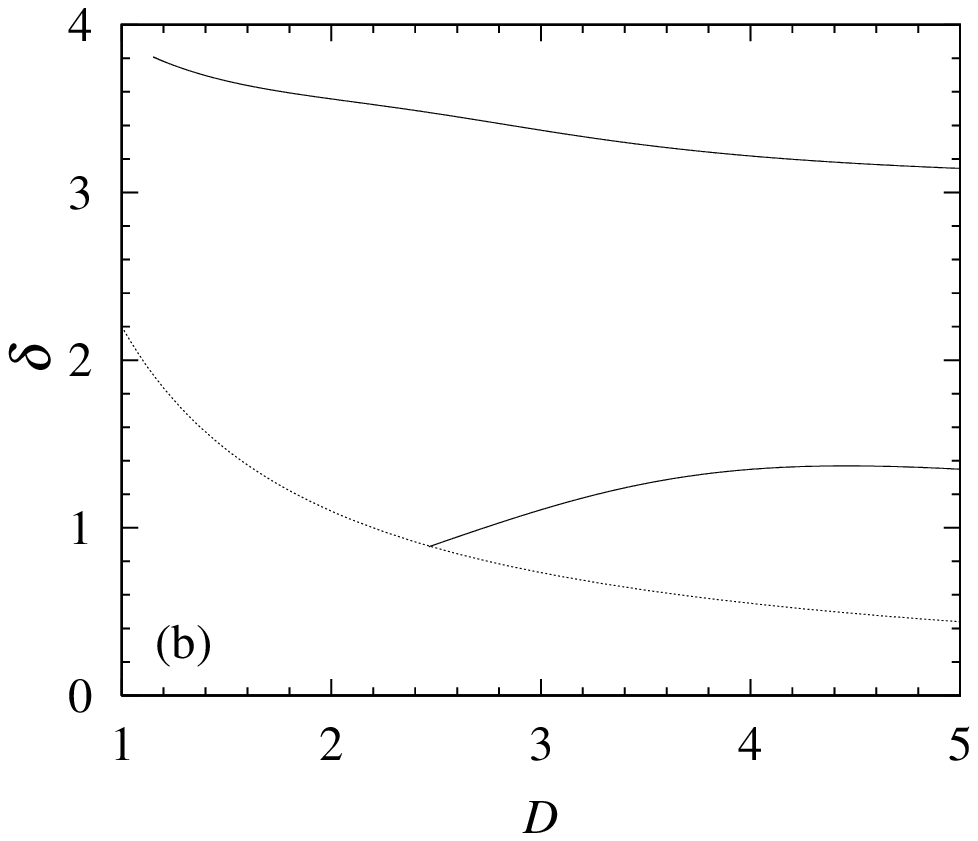}
\end{center}
\caption{Branches of the discrete ES solutions for $k=0.3$ and
$\protect\gamma _{1}=\protect\gamma _{2}=0.05$. (a) Solutions in
the $(\protect\delta ,q)$-plane for $(D,N)=(100,86)$ (solid line)
$(10,31)$ (dashed line) and $(5,23)$ (broken line). According to
(\protect\ref{area_scale}) ESs also exist only above the dotted
line $q=0.6$ for this $k$-value. Labeled points correspond to the
relevant subpanels of Fig.~\protect\ref{fig:4b} where solutions
profiles are displayed. (b) Solutions in the $(D,\protect\delta
)$-plane for $q=5$ and $N=22$. Note that, according to
(\protect\ref{area_scale}), we must have $\protect\delta D>2.2$
(the boundary is shown as a dotted line). } \label{fig:4a}
\end{figure}

\begin{figure}[tbp]
\begin{center}
\includegraphics[scale=0.55]{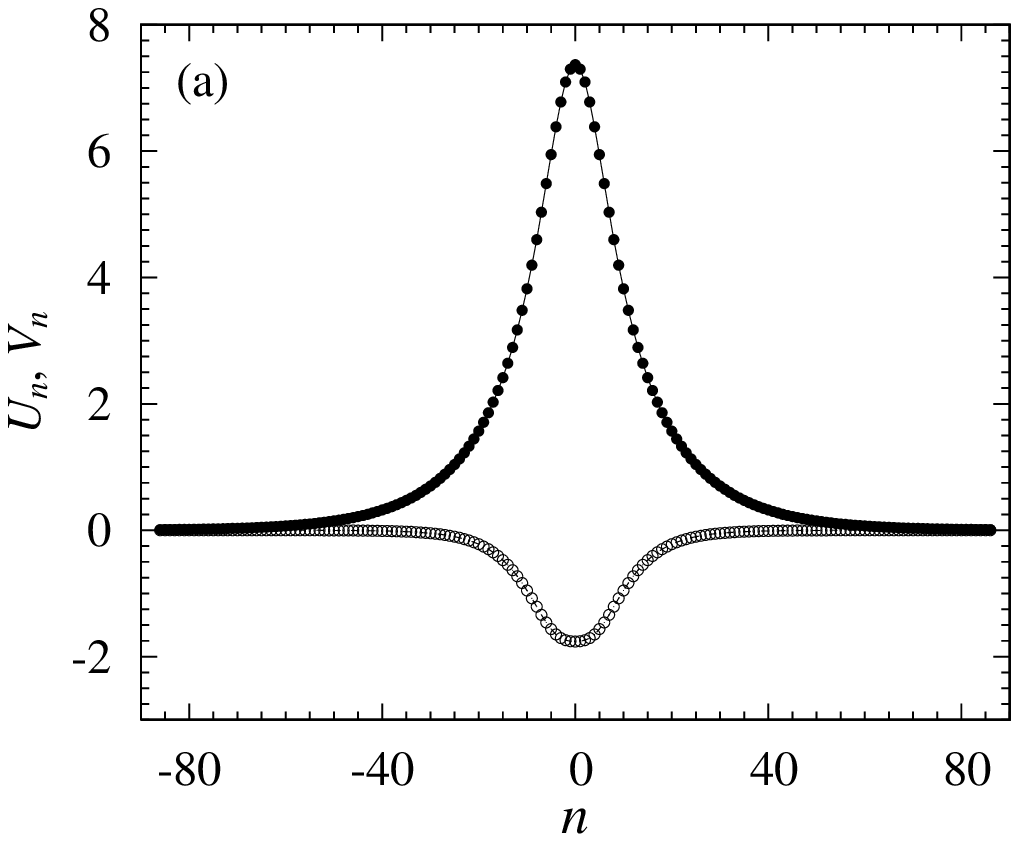}\qquad \includegraphics[scale=0.55]{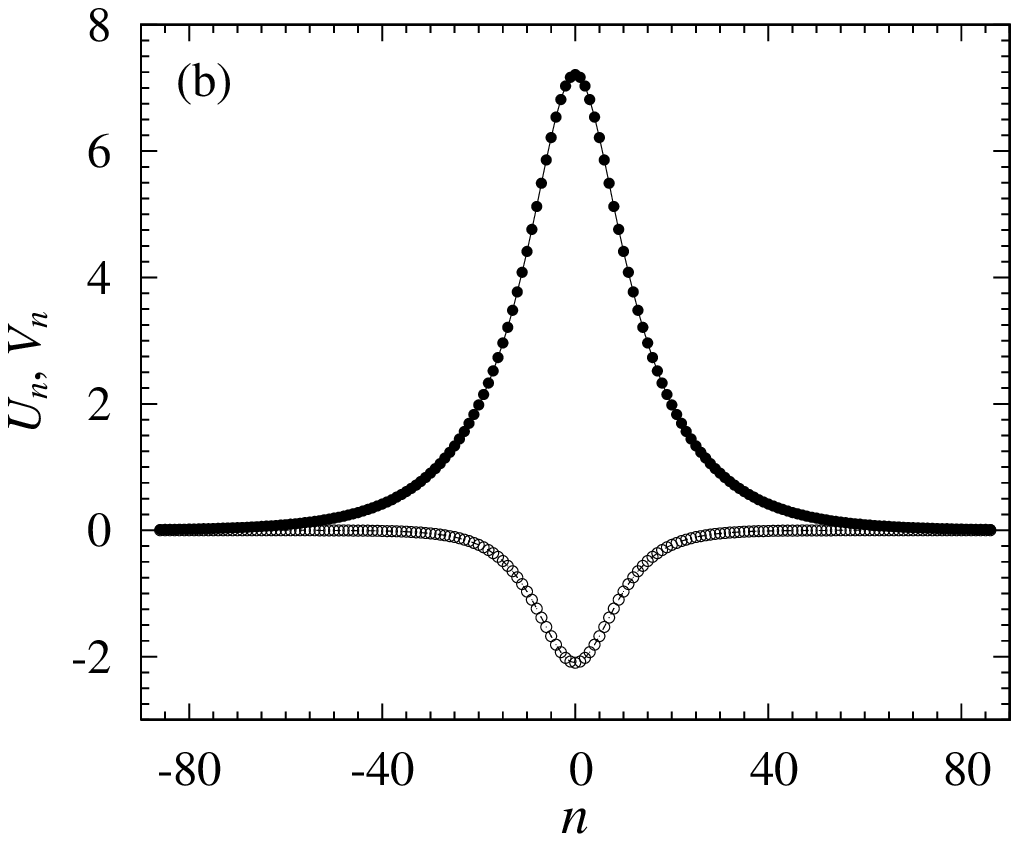}\\[2ex]
\includegraphics[scale=0.55]{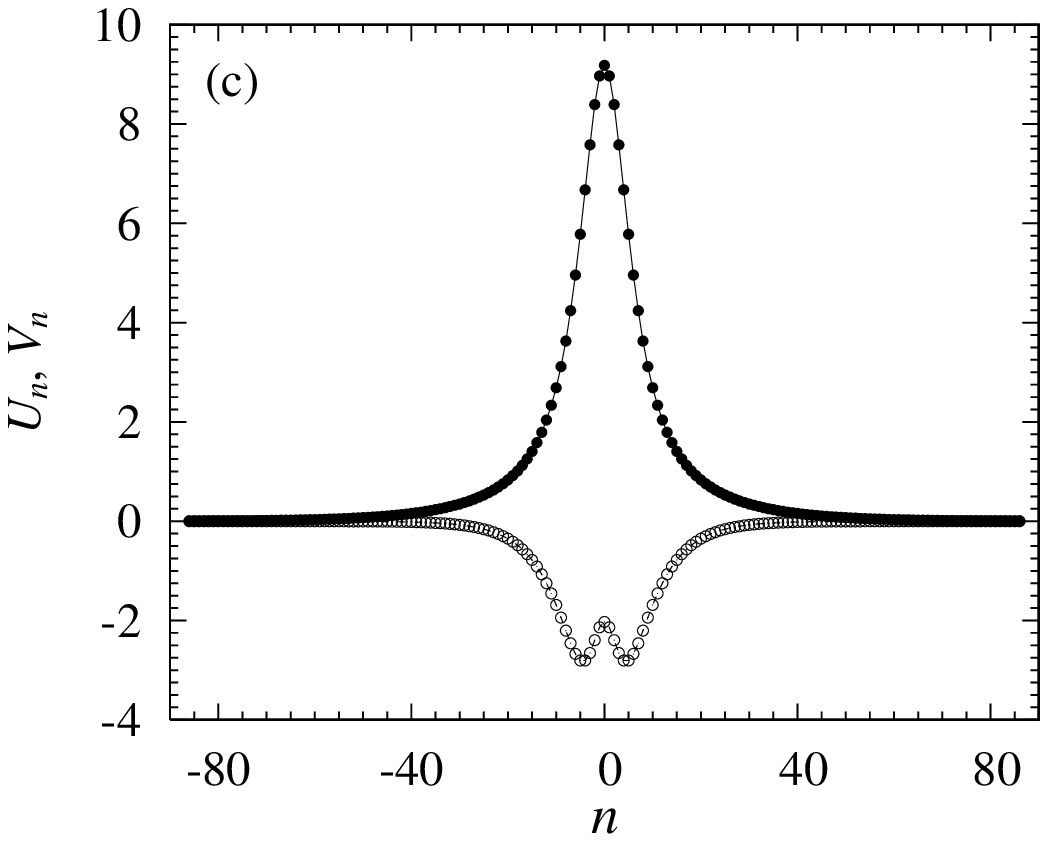}\qquad \includegraphics[scale=0.55]{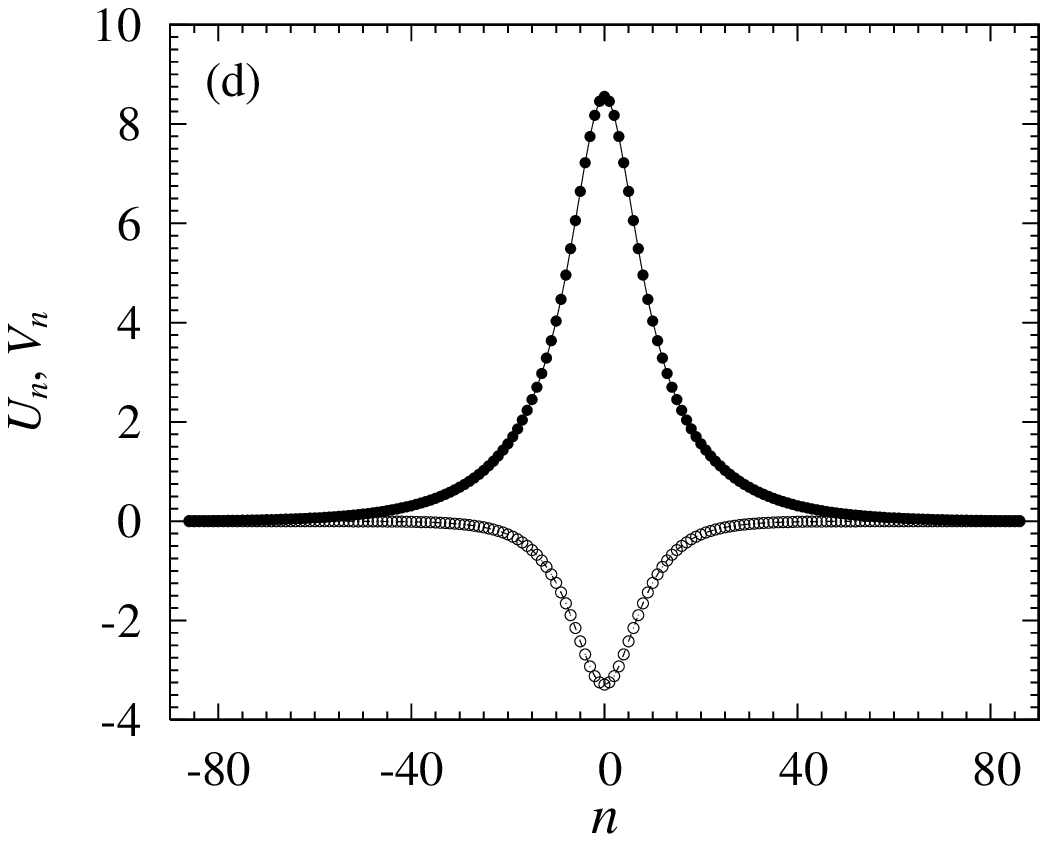}\\[2ex]
\includegraphics[scale=0.55]{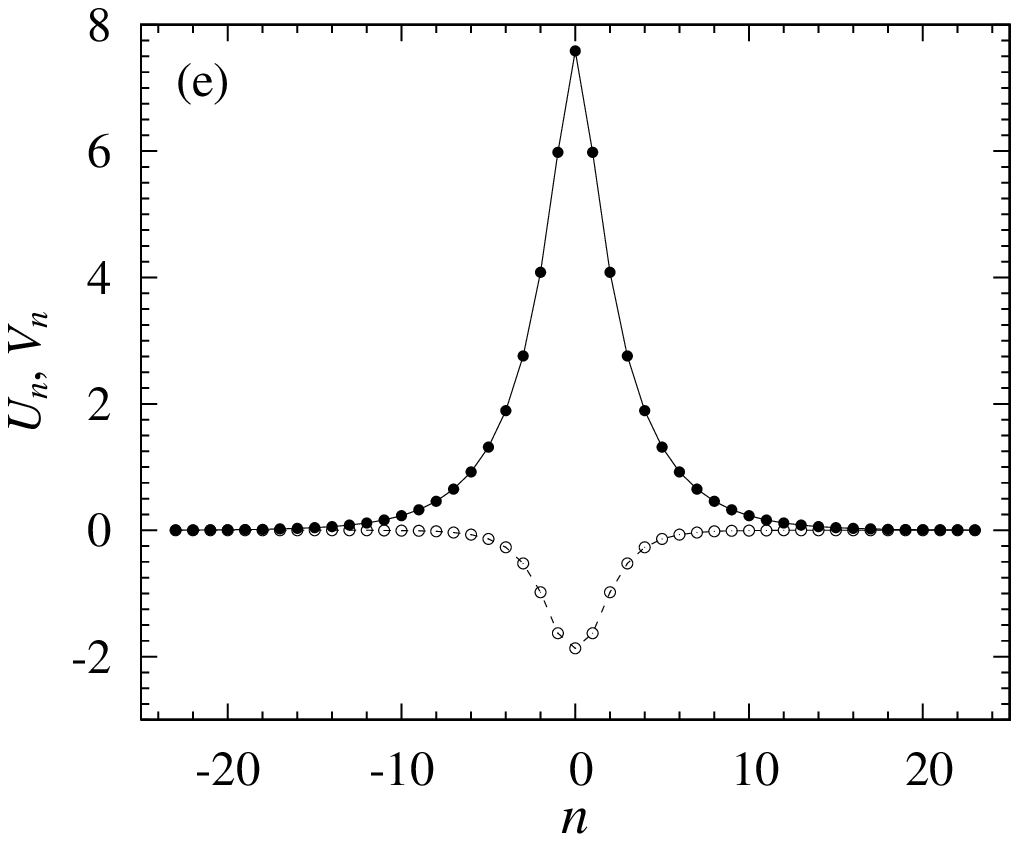}\qquad \includegraphics[scale=0.55]{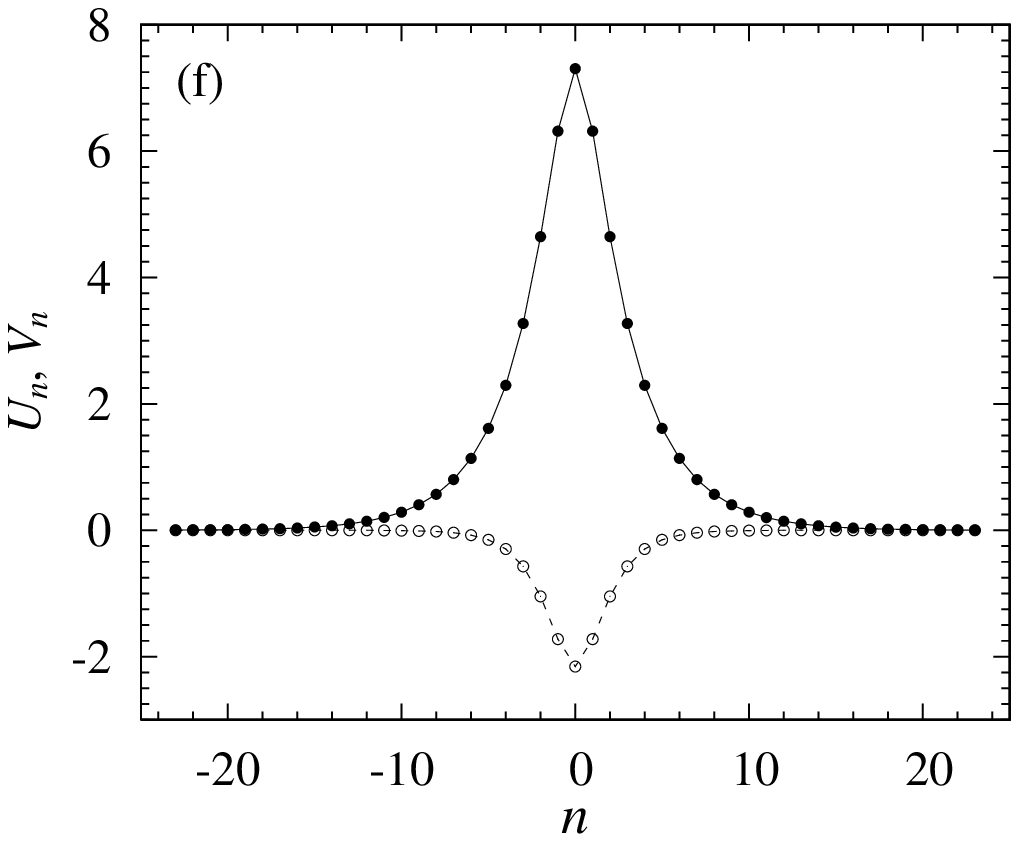}\\[2ex]
\includegraphics[scale=0.55]{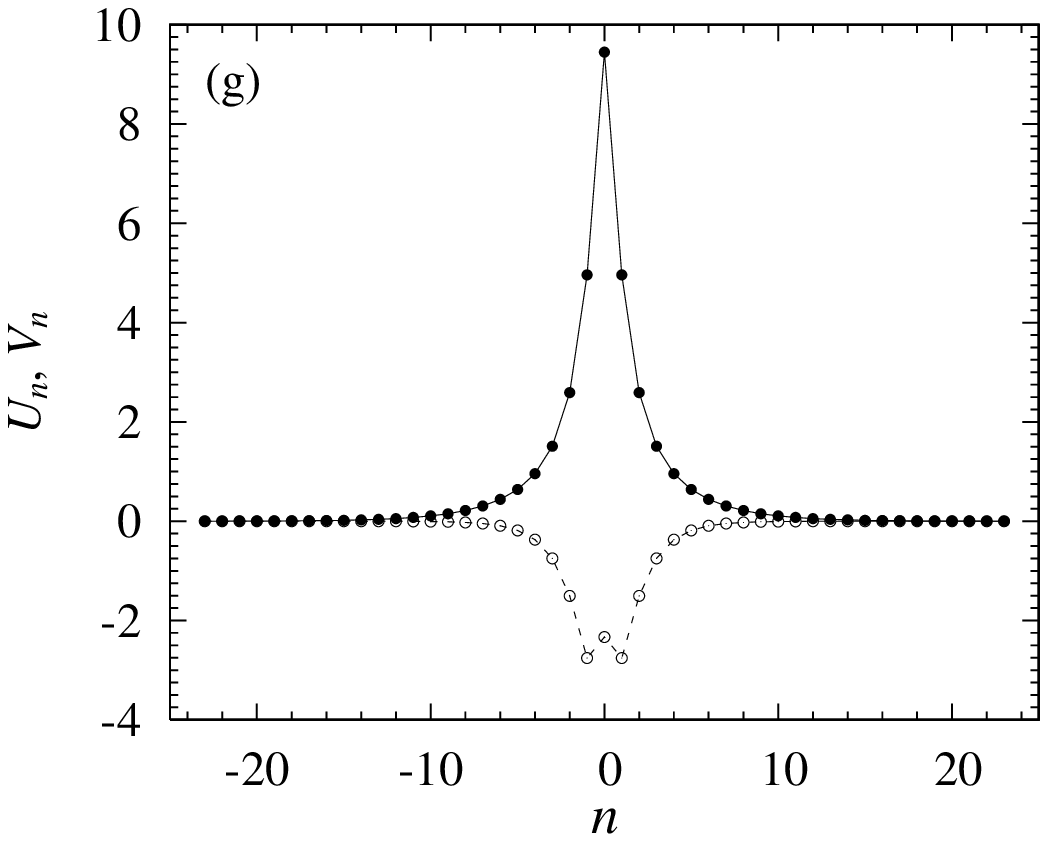}\qquad \includegraphics[scale=0.55]{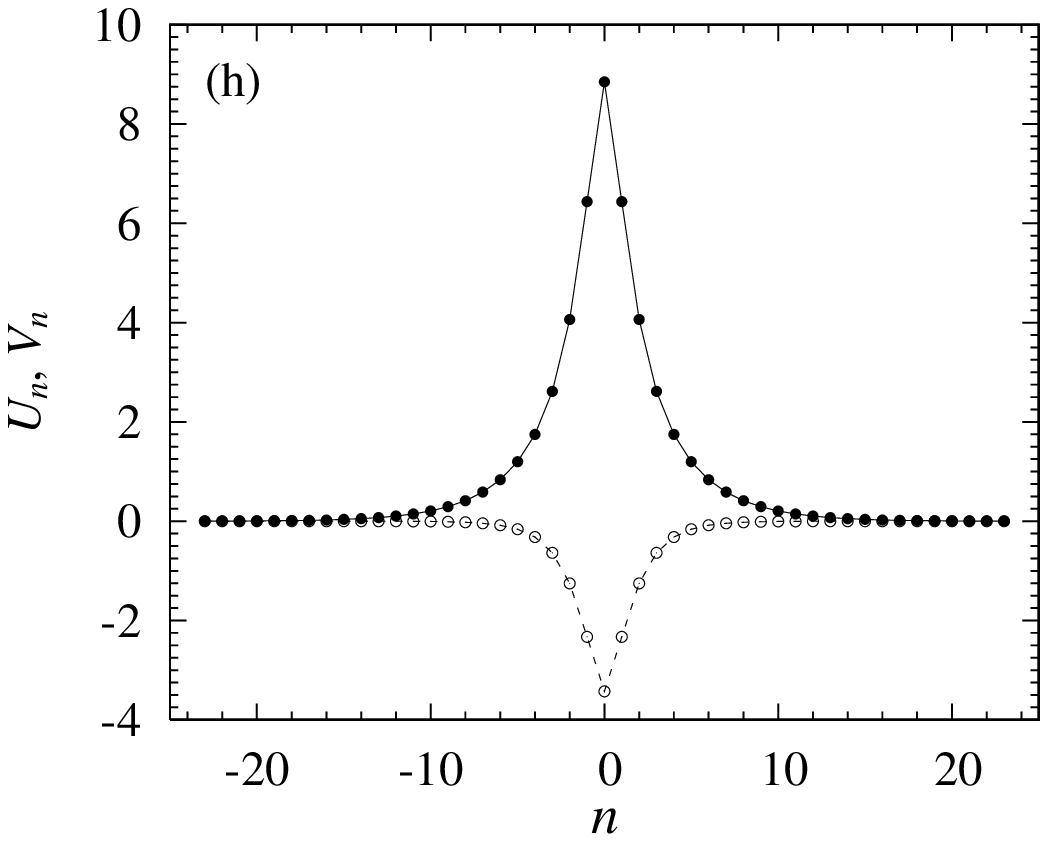}
\end{center}
\caption{Profiles of the discrete ESs corresponding to the labeled
points in Fig.~\protect\ref{fig:4a}(a): (a) $\protect\delta
=0.58637 $ and $q=5$; (b) $\protect\delta =2.9894$ and $q=5$; (c)
$\protect\delta =0.65467$ and $q=0.6$; (d) $\protect\delta
=3.3693$ and $q=0.6$; (e) $\protect\delta =1.3496 $ and $q=5$; (f)
$\protect\delta =3.1432$ and $q=5$; (g) $\protect\delta =0.80213$
and $q=0.6$; (h) $\protect\delta =3.3824$ and $q=0.6$. In
panels~(a)-(d), $D=100$ and $N=85$, and in panels (e)-(h), $D=5$
and $N=22$. In this and all subsequent plots, the FF
($u$-component) is interpolated by a solid line and the SH
($v$-component) by a broken line. } \label{fig:4b}
\end{figure}

Figure~\ref{fig:4a} depicts branches of discrete ESs in the
presence of $\chi ^{(3)}$ terms, with $\gamma
_{1}=\gamma_{2}=0.05$ and $k=0.3$. Two distinct branches of ESs
are displayed. Along each branch, the profile changes
continuously. Figure~\ref{fig:4b} displays an array of profiles of
these ESs for $D=100$ and $D=5$. Note that the second
(higher-$\delta$) branch can be considered to be a family of
\emph{fundamental}, i.e.\ single-humped, solitons throughout the
range of existence, whereas the double-humped structure of the SH
component of the first branch becomes evident for small $q$. This
property, and the location of the branches in the $(\delta ,q)$
plane, are fully consistent with results obtained in the continuum
limit \cite[Fig.~3]{ChMaYaKa:01}. Our numerical results also
suggest that there may exist further branches for lower $\delta
$-values, each subsequent one containing more oscillations in core
of the SH component. However, computation becomes unreliable
beyond the first two branches for the reasons given at the end of
the last section, and so we do not display those results here.
Also experience from the continuum model suggests that
non-fundamental ESs never have a chance to be stable.

Figure \ref{fig:4a}(b) depicts continuation in $D$ of the solutions on these
two branches for $q=5$. In this case, we find that there is a lower limit on
$D$ beyond which no ESs exist. This corresponds to the right-hand inequality
in the first condition in (\ref{area_scale}).

\begin{figure}[tbp]
\begin{center}
\includegraphics[scale=0.7]{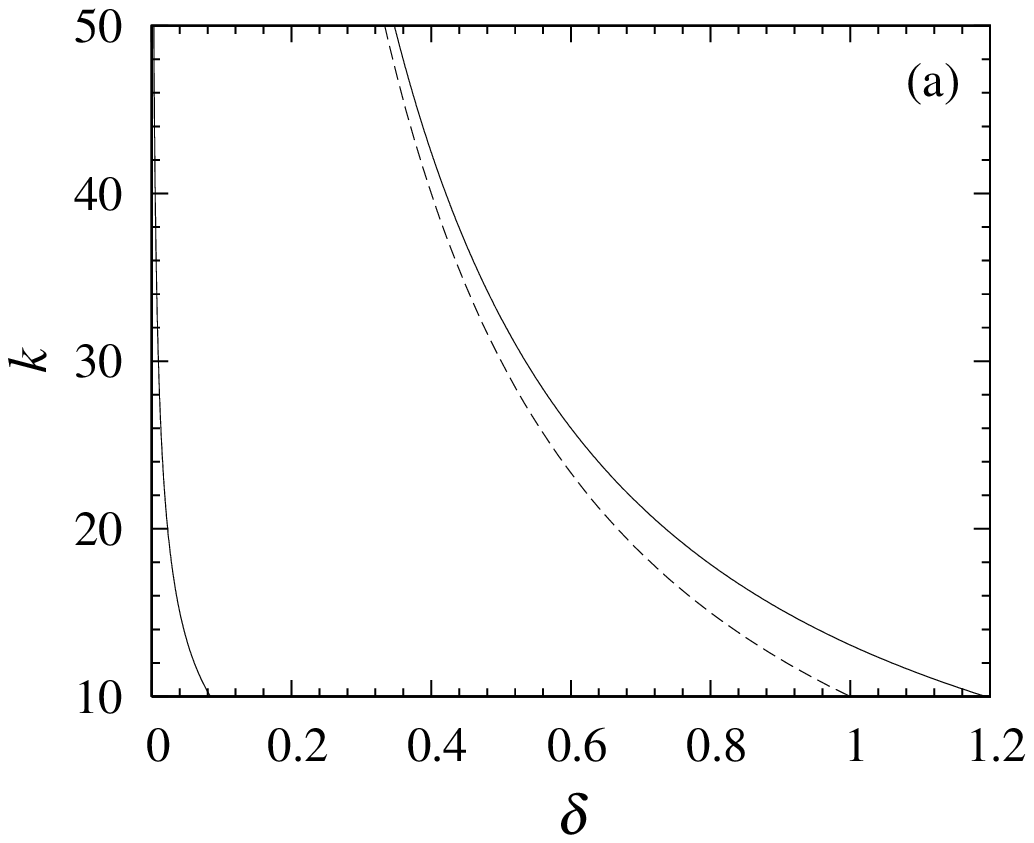}\quad \includegraphics[scale=1.23]{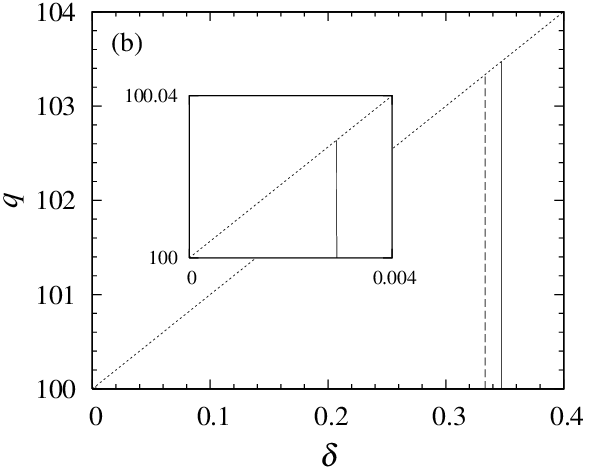}
\end{center}
\caption{Branches of discrete ESs for $\protect\gamma
_{1}=\protect\gamma_{2}=0.05$ and $D=10$; $N=7$ is chosen for
larger $\protect\delta$, and $N=5$ for smaller $\protect\delta $.
The analytical prediction given by Eqs.~(\protect\ref{final}) and
(\protect\ref{rescaledpars}) is plotted as a dashed line. (a)
Solutions in the $(\protect\delta,k)$-plane for $q=2k$. (b)
Solutions in the $(\protect\delta,q)$-plane for $k=50$. The ESs
exist in the region $100<q<100+10\protect\delta $, whose boundary
is shown as a dotted line.} \label{fig:4b1}
\end{figure}

\begin{figure}[tbp]
\begin{center}
\includegraphics[scale=0.55]{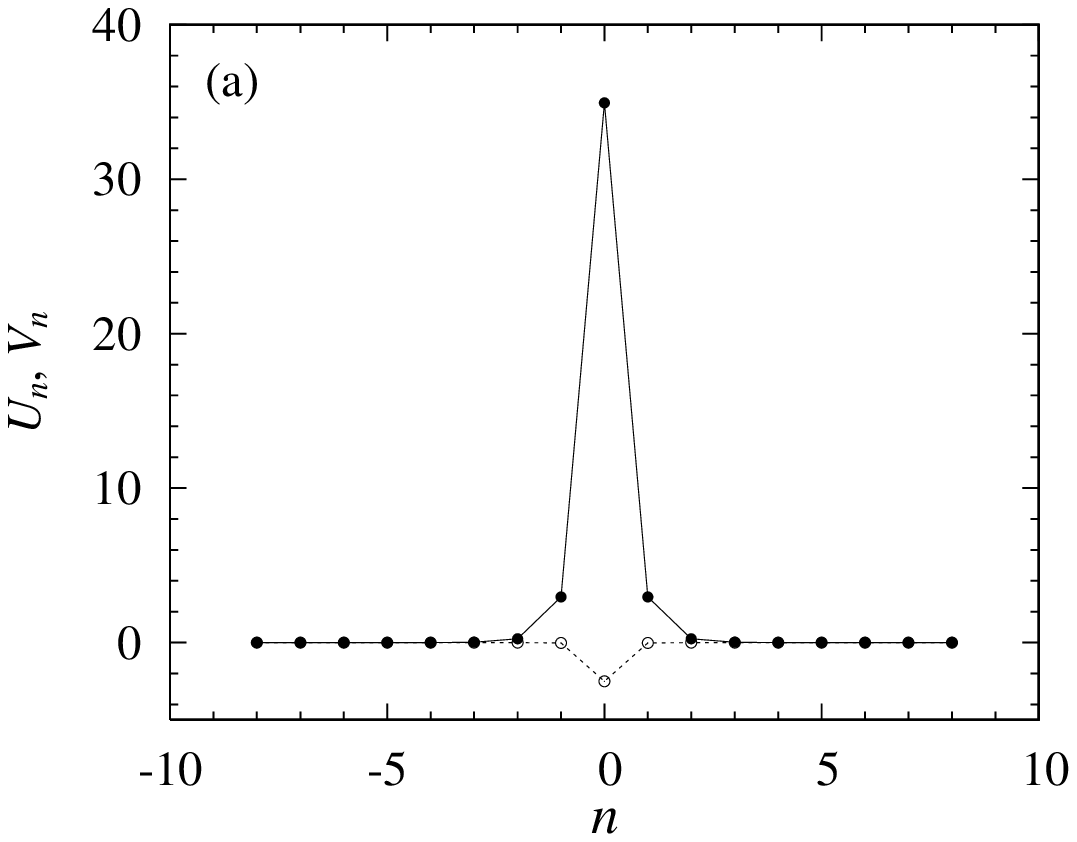}\qquad \includegraphics[scale=0.55]{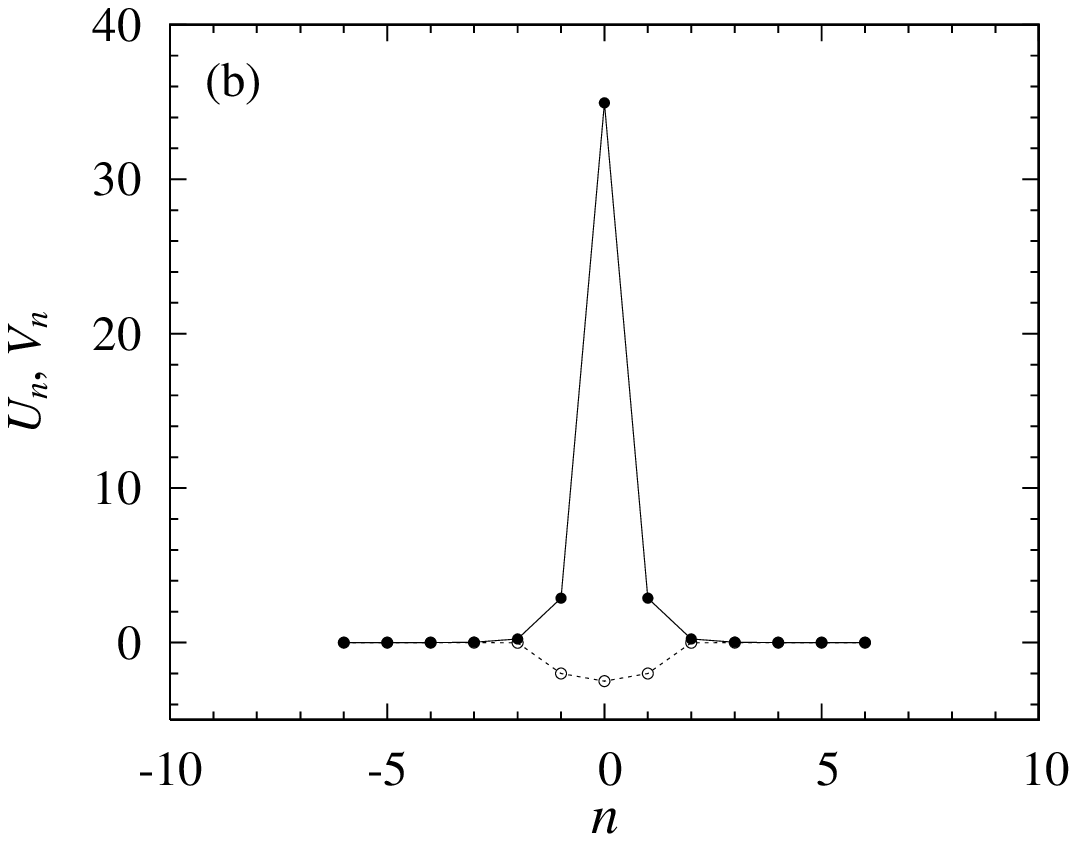}
\end{center}
\caption{Examples of the ESs on solution branches in
Fig.~\protect\ref{fig:4b1}(b): (a) $\protect\delta =0.34718$,
$q=102$ and $N=7$; (b) $\protect\delta =0.0029014$, $q=100.02$ and
$N=5$. } \label{fig:4b2}
\end{figure}

We will now check the validity of the analytical approximation
developed in Section~3 when $\tilde{k}$ and hence $k$ are large.
In Fig.~\ref{fig:4b1}, we display the two numerically computed ES
branches with the same values of $\gamma _{1}$ and $\gamma _{2}$
as those in Fig.~\ref{fig:4a} for $D=10$ when $k$ is rather large,
and compare them with the analytical prediction (\ref{final}),
taking into regard the relations (\ref{rescaledpars}). In
Fig.~\ref{fig:4b1}(a), the parameters $\delta $ and $k$ are varied
with $q=2k$, and in Fig.~\ref{fig:4b1}(b) the parameters $\delta $
and $q$ are varied for $k=50$. The analytical predictions are
plotted as dashed lines. We find good agreement between the
numerical and analytical results, especially for large $k$.
Figure~\ref{fig:4b2} displays the profiles of these ESs. The
fundamental soliton in Fig.~\ref{fig:4b2}(a) has a steep peak at
$n=0$, as assumed in the approximate analysis of Section~3.

\begin{figure}[tbp]
\begin{center}
\includegraphics[scale=0.55]{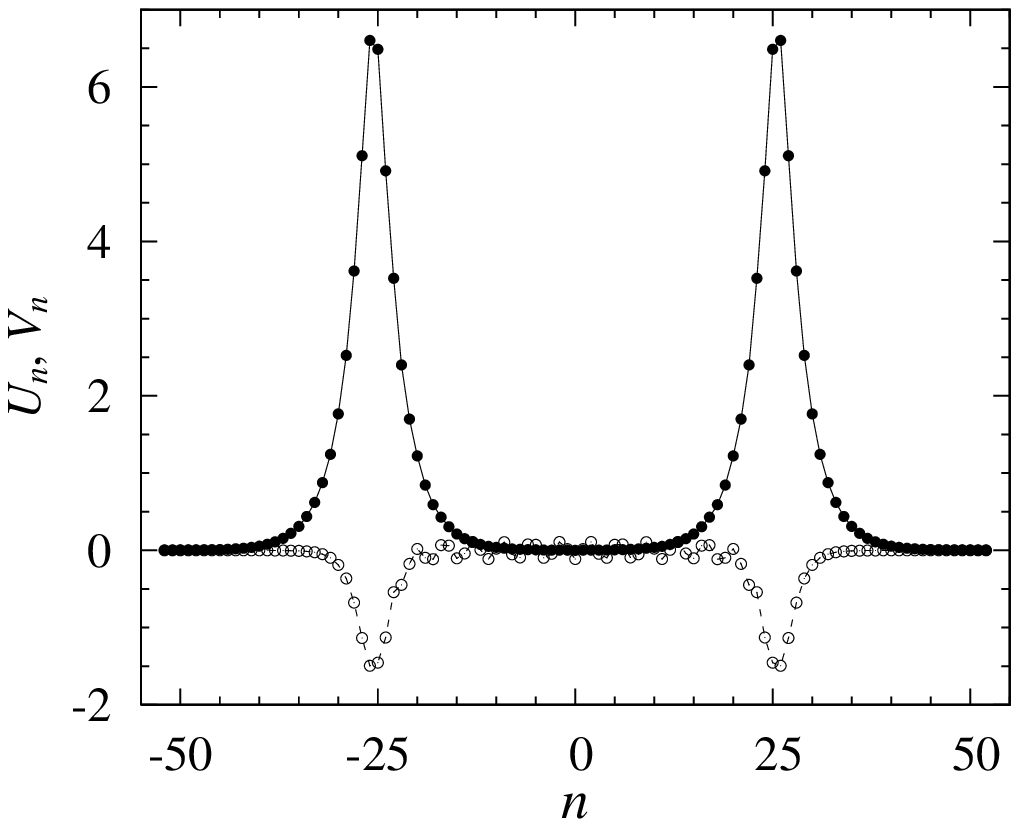}
\end{center}
\caption{A typical example of a two-pulse bound state ES in the
lattice system (\protect\ref{eqn:disc}) with
$\protect\delta=1.1261$, $\protect\gamma _{1}=\protect\gamma
_{2}=0.05$, $D=5$, $k=0.3$, $q=7$ and $N=51$.} \label{bound-state}
\end{figure}

It is well known that in the continuum case multi-pulse homoclinic
orbits exist (under a mild \emph{Birkhoff signature} condition
\cite{MiHoOR:92,KoChBuSa:02}) along families of curves in the
parameter plane that accumulate on the curves of the fundamental
ES solutions. Unlike the branches computed above, they do not
feature a single-humped shaped in either component, but rather
look like bound-states of two spatially separated fundamental
solitons. We have found exactly the same solution families in the
discrete model too. An example is presented in
Fig.~\ref{bound-state}.

\subsection{The case of\/ $\protect\delta <0$}

\begin{figure}[tbp]
\begin{center}
\includegraphics[scale=0.7]{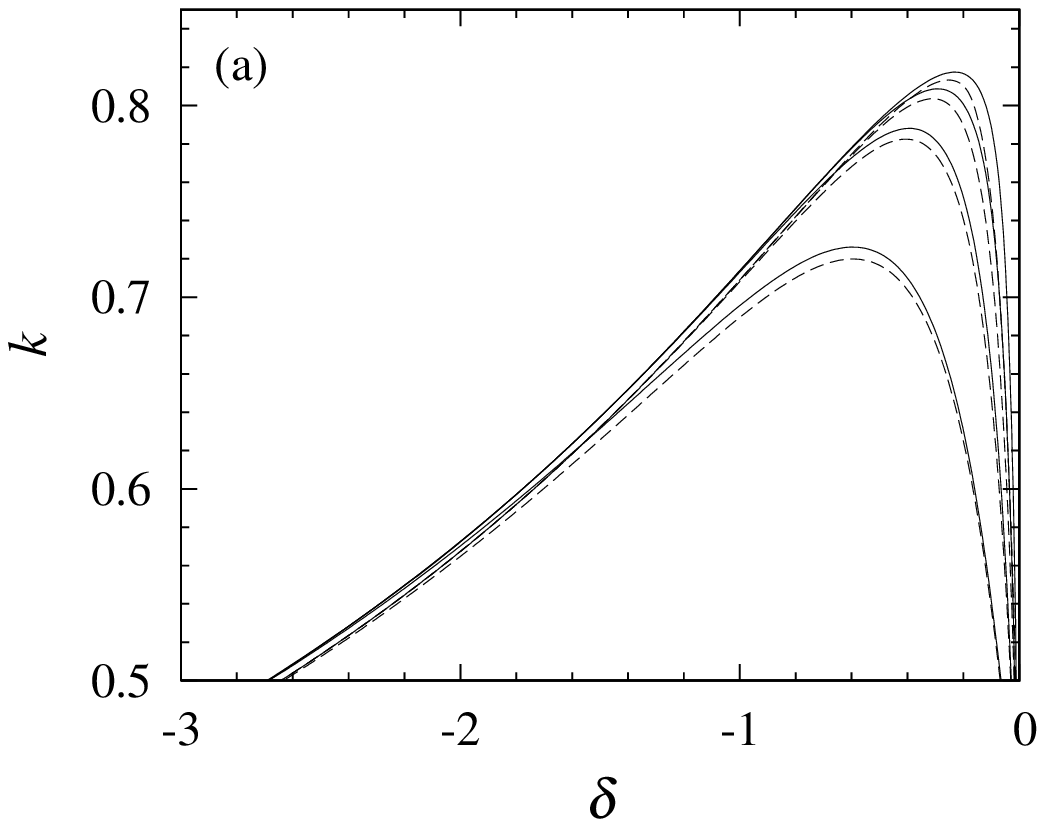}\quad \includegraphics[scale=0.7]{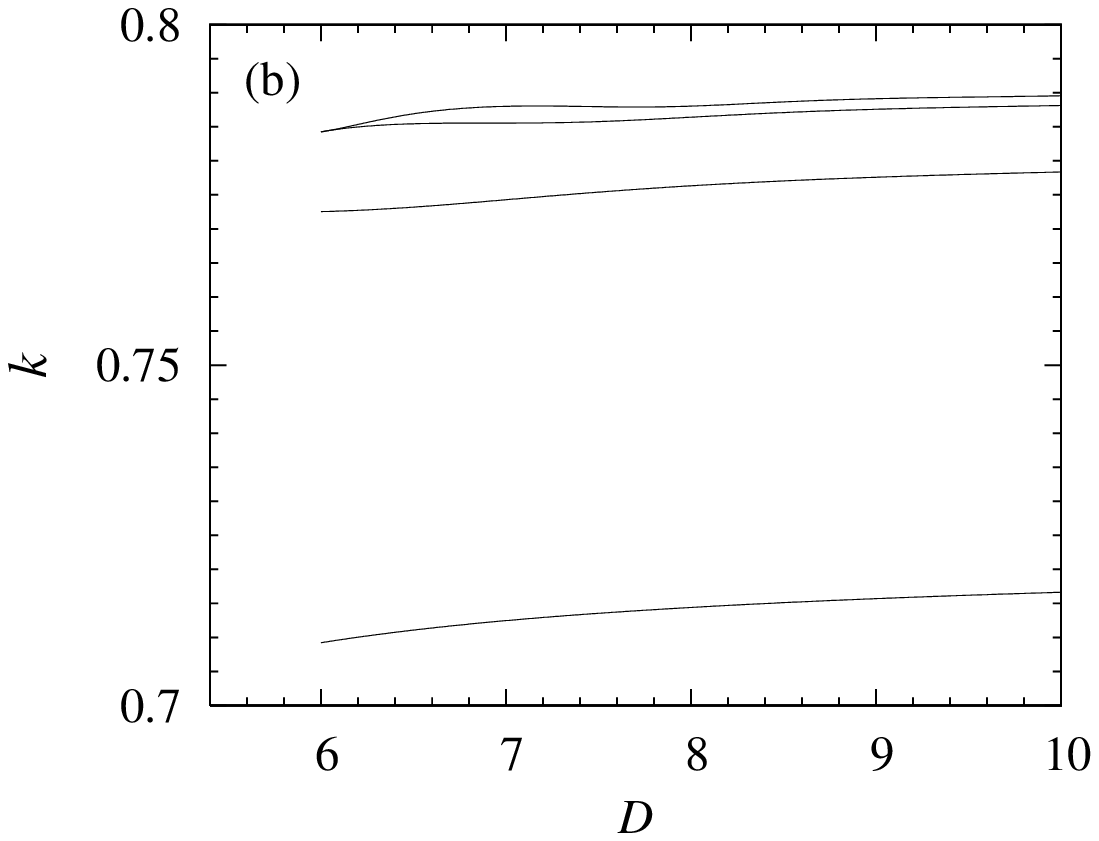}
\end{center}
\caption{Branches of ESs in the discrete system
(\protect\ref{eqn:lattice}) with $q=1$ and $\protect\gamma
_{1}=\protect\gamma _{2}=-0.05$: (a) In the $(\protect\delta
,k)$-plane for $D=100$ and $10$; (b) in the $(D,k)$-plane for
$\protect\delta =-0.5$ and $N=30$. In panel~(a), the solid and
dashed curves represent the results for $D=100$ and $10$,
respectively. When $D=100$ (resp. $D=10$), $N=85$ (resp. $N=30$)
were used for the most part but $N=93$ or $N=109$ (resp. $N=35$ or
$N=40$) were used for the two outer curves with large
$|\protect\delta |$. According to condition
(\protect\ref{area_scale}), ESs exist only above $k=0.5$ and below
$k=0.5-\protect\delta D$, when $q=1 $. } \label{fig:4c}
\end{figure}

\begin{figure}[tbp]
\begin{center}
\includegraphics[scale=0.55]{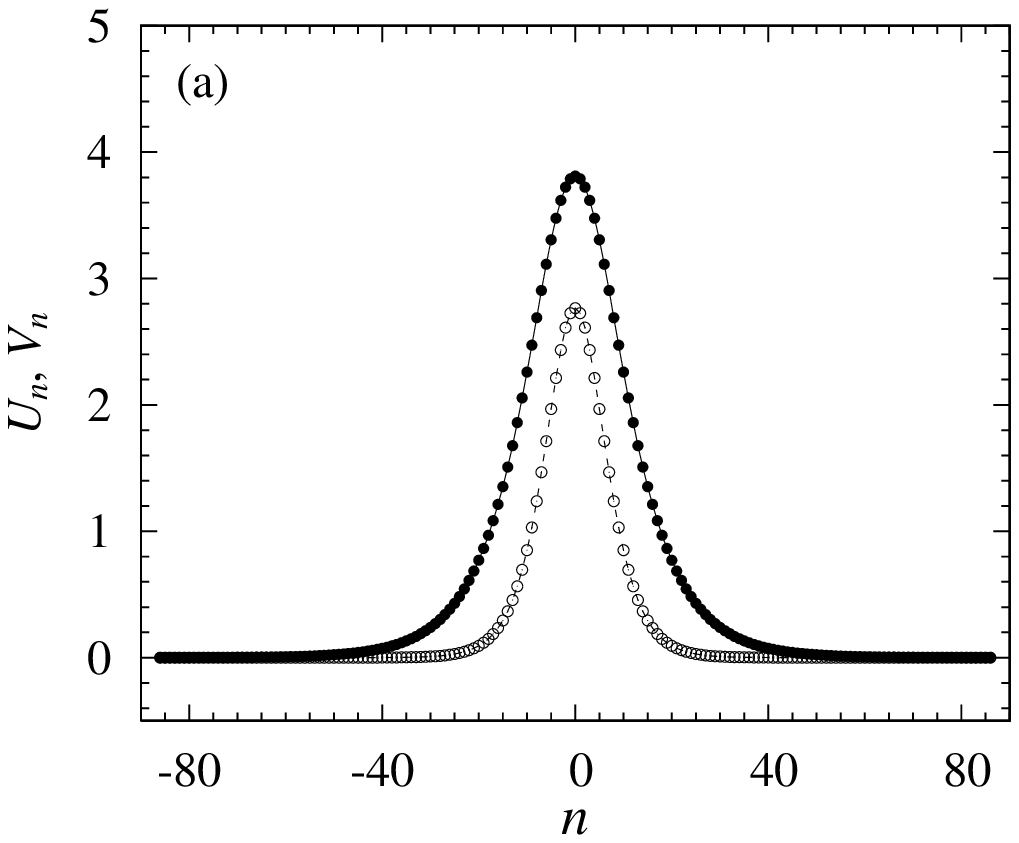}\qquad \includegraphics[scale=0.55]{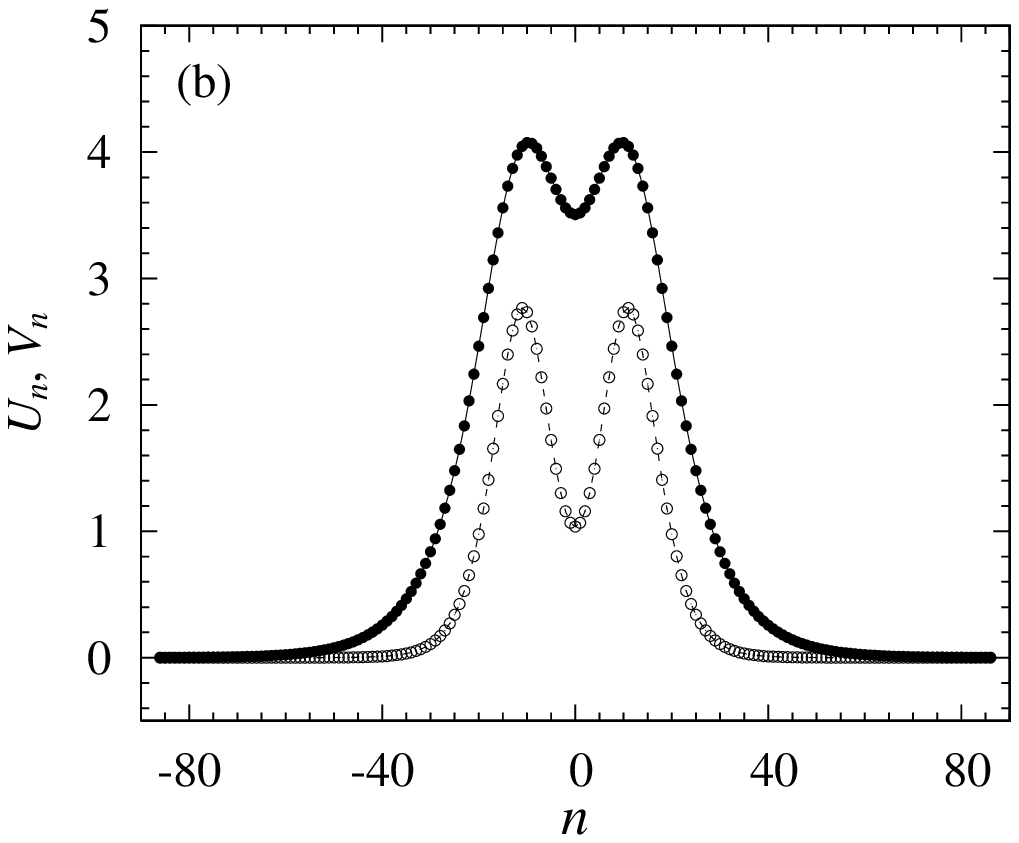}\\[2ex]
\includegraphics[scale=0.55]{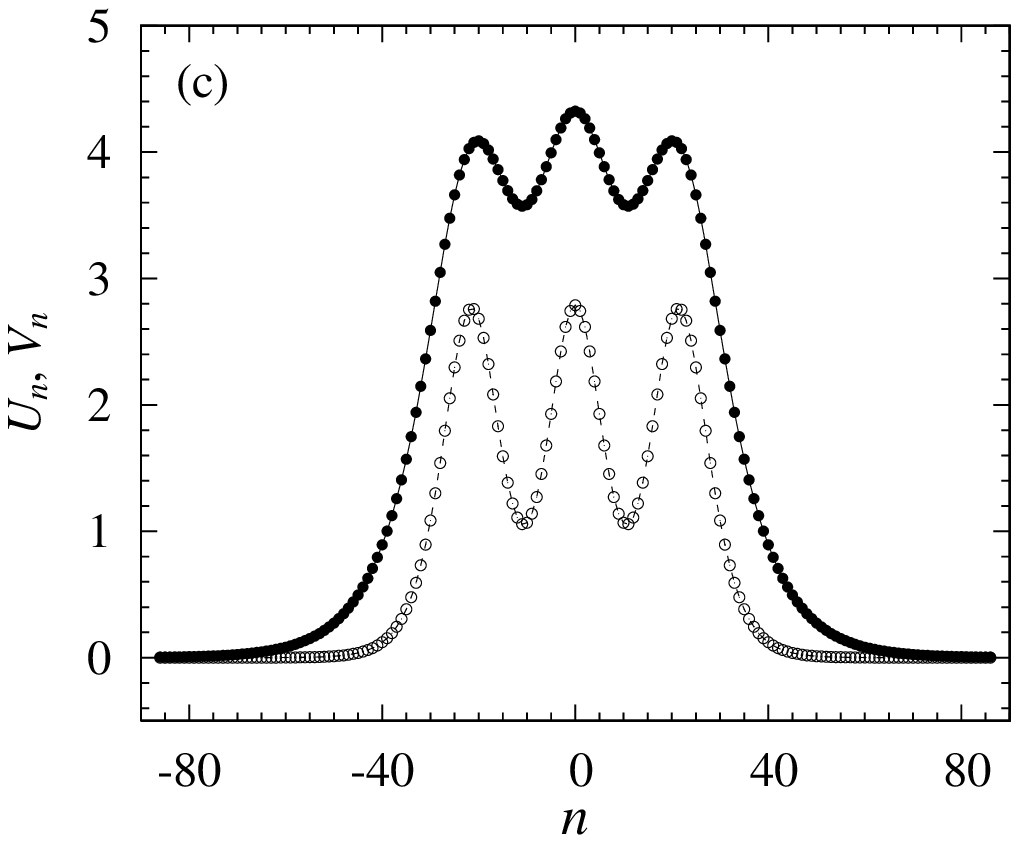}\qquad \includegraphics[scale=0.55]{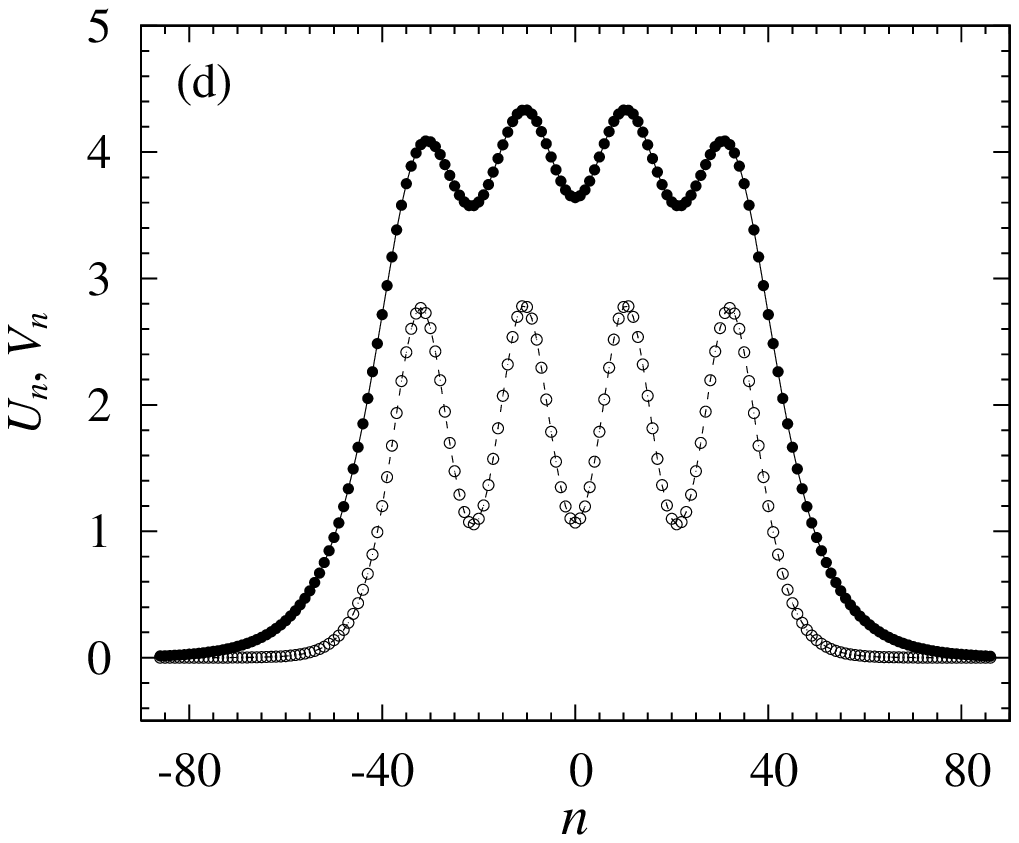}\\[2ex]
\includegraphics[scale=0.55]{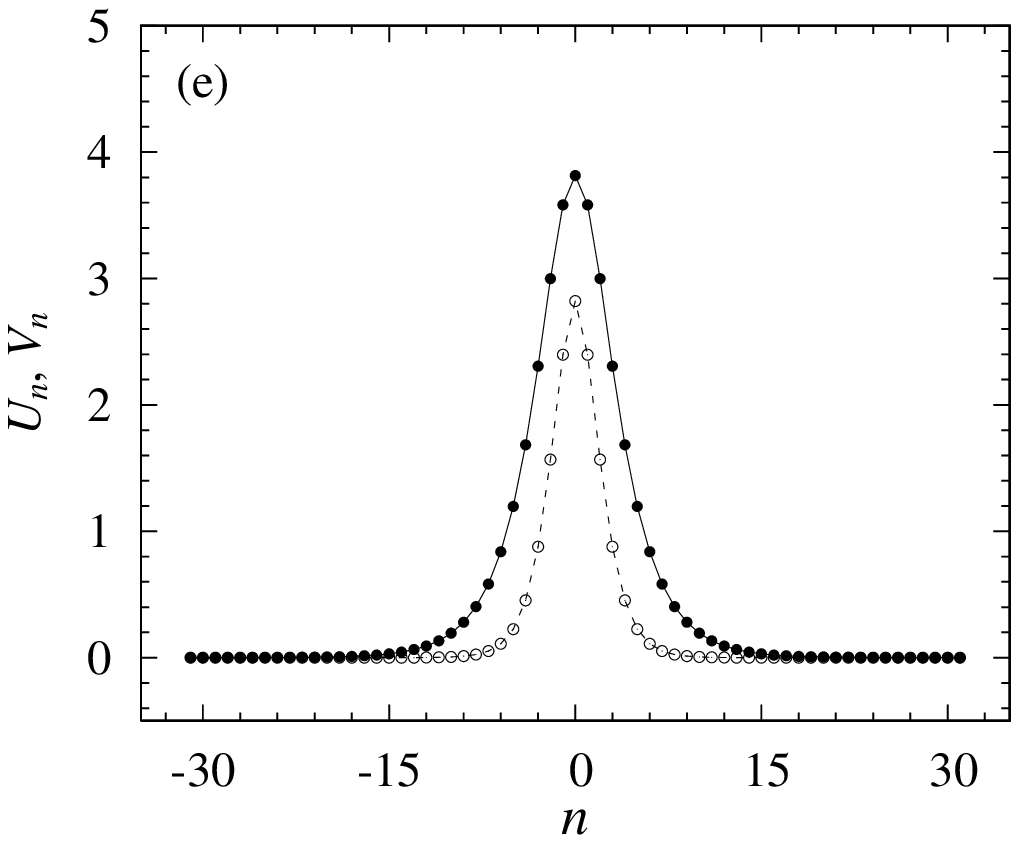}\qquad \includegraphics[scale=0.55]{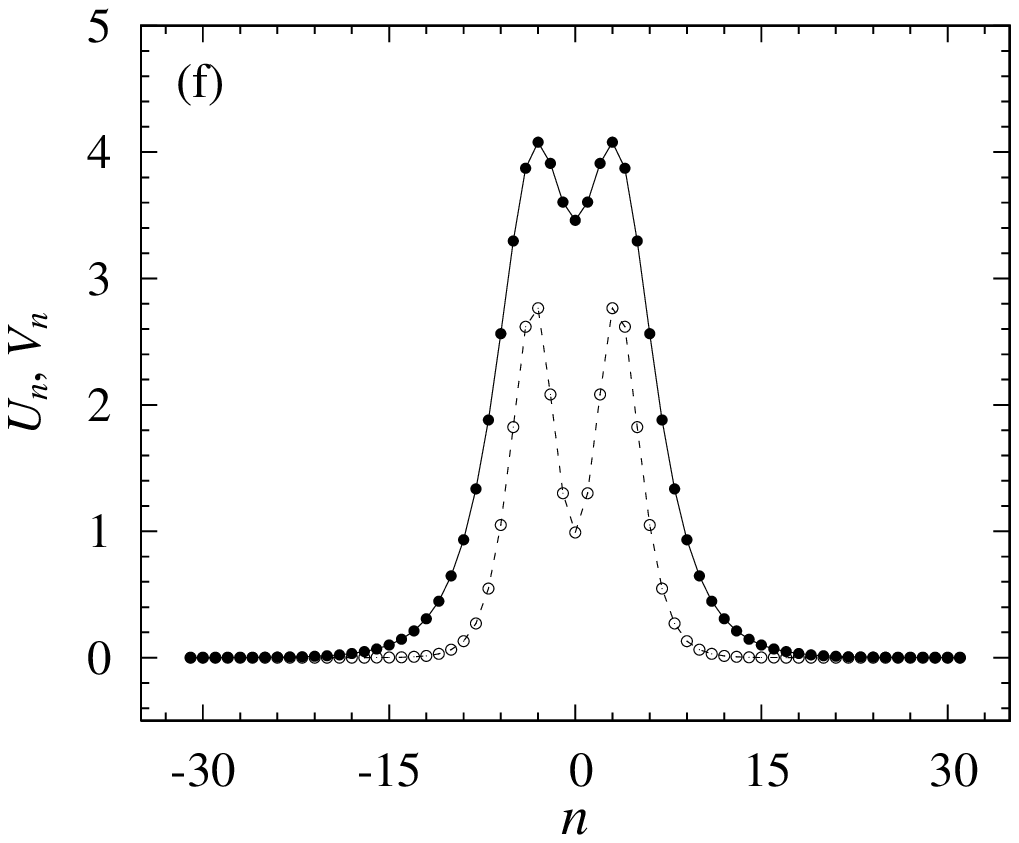}\\[2ex]
\includegraphics[scale=0.55]{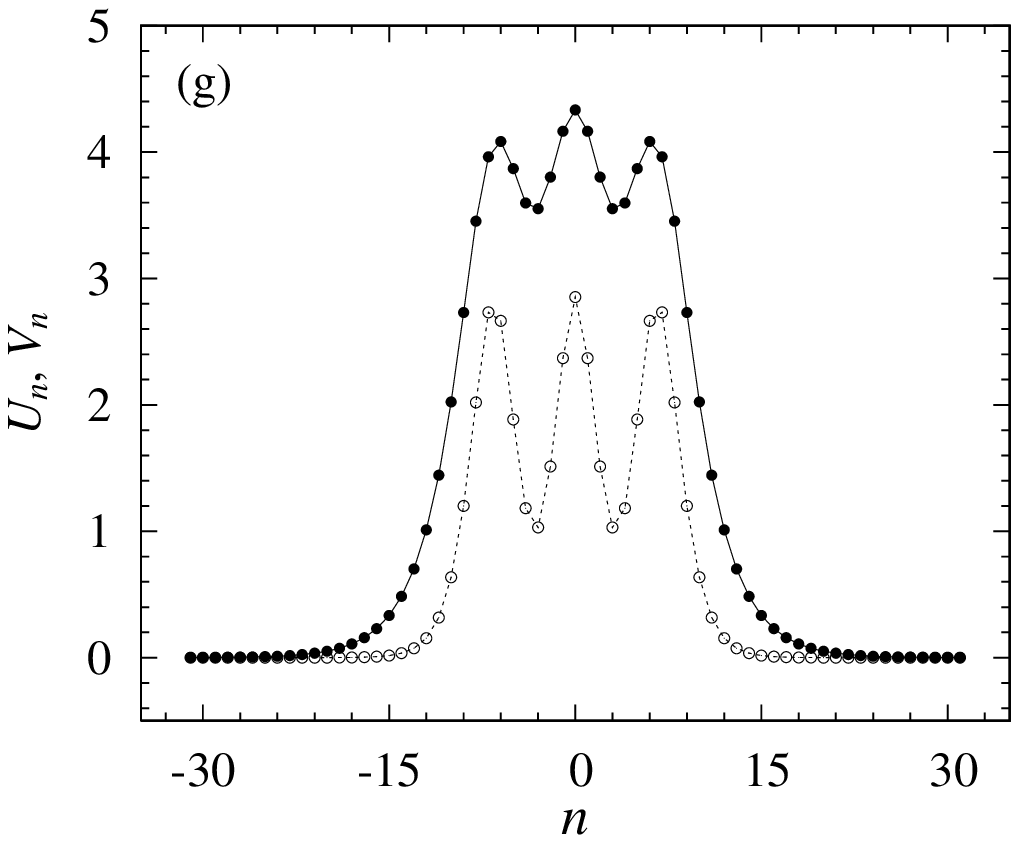}\qquad \includegraphics[scale=0.55]{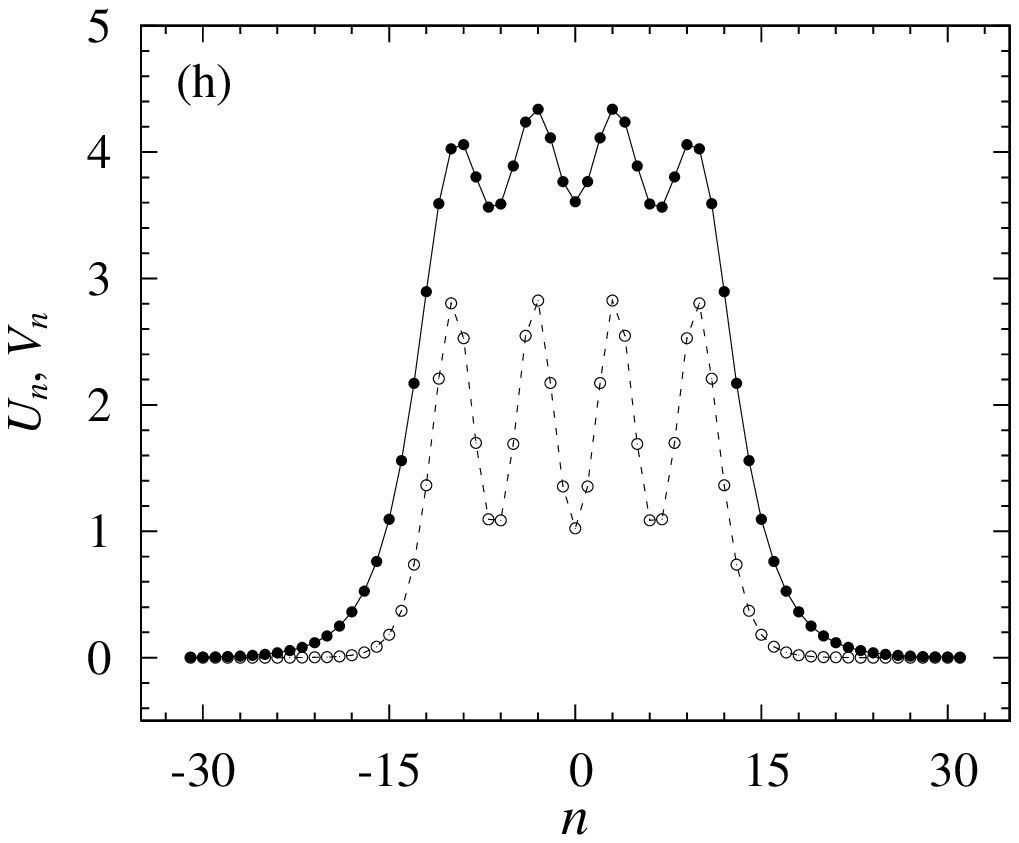}
\end{center}
\caption{ESs on the branches in Fig.~\protect\ref{fig:4c}(a) for
$\protect\delta =-1$: (a) $k=0.69564$; (b) $k=0.71312$; (c)
$k=0.71383$; (d) $k=0.71387$; (e) $k=0.6895$; (f) $k=0.70811$; (g)
$k=0.70898$; (h) $k=0.70902$. In panels (a)-(d), $D=100$ and
$N=85$, and in panels~(e)-(h), $D=10$ and $N=30$. } \label{fig:4d}
\end{figure}

\begin{figure}[tbp]
\begin{center}
\includegraphics[scale=0.55]{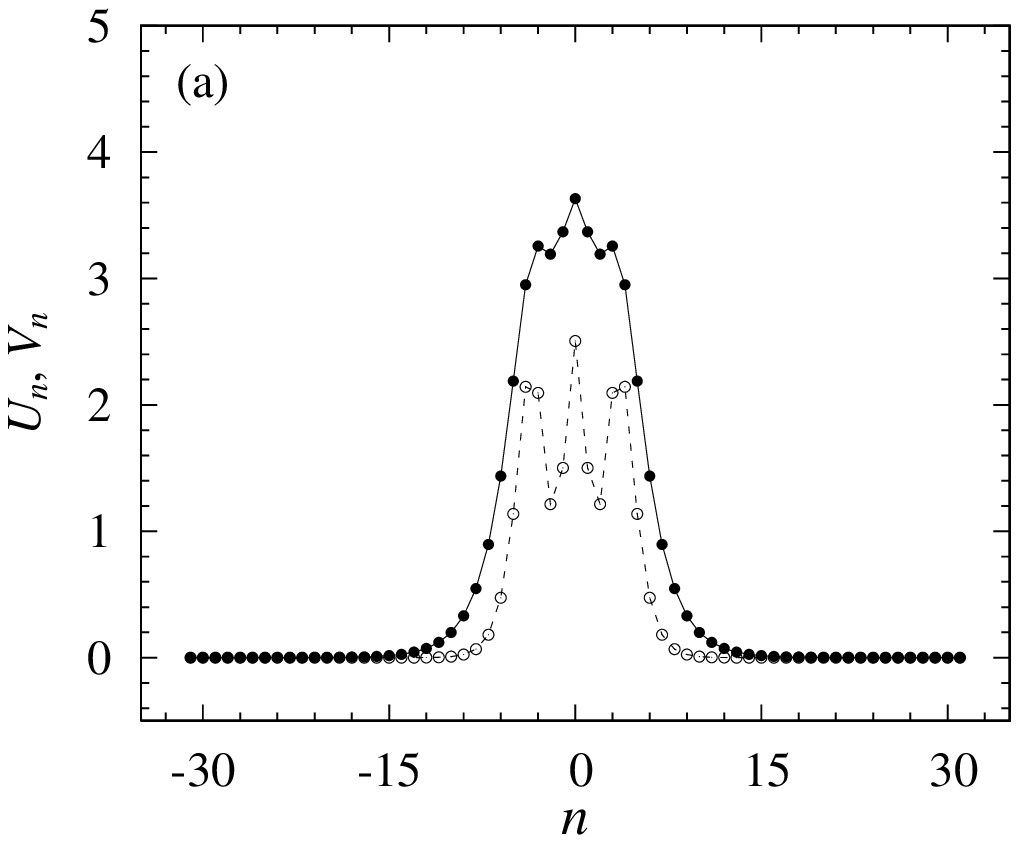}\quad \includegraphics[scale=0.55]{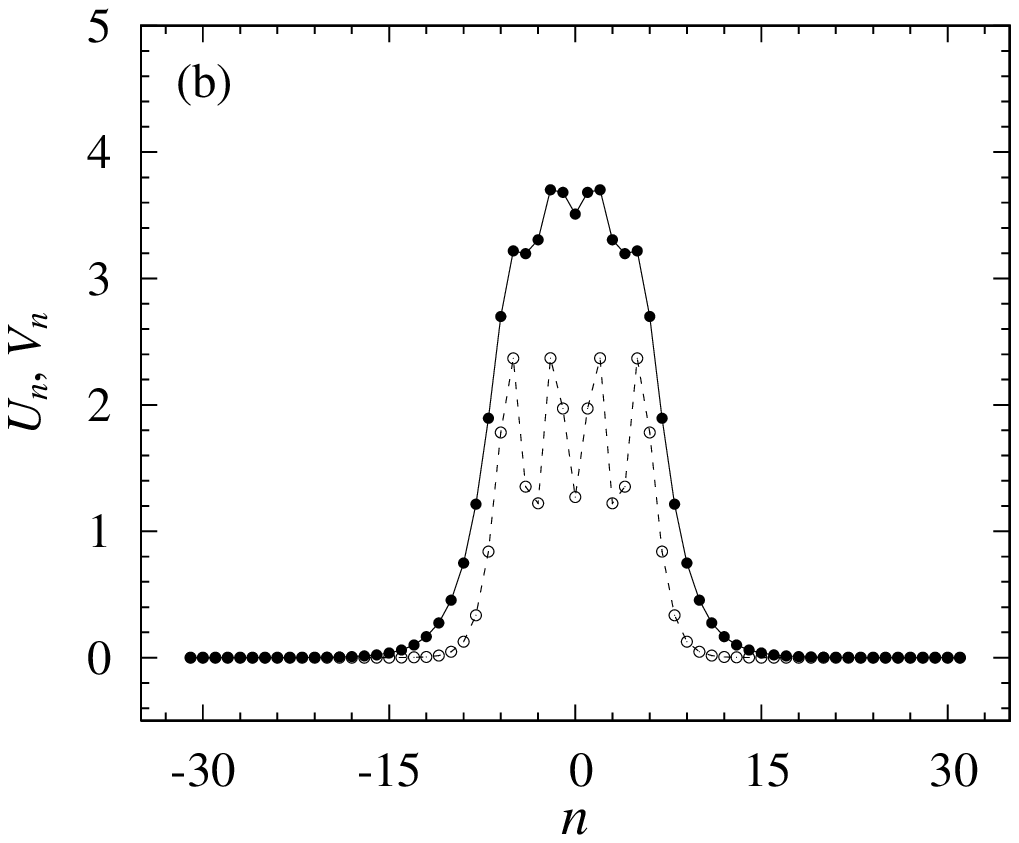}
\end{center}
\caption{ESs at the end of the above two branches in
Fig.~\protect\ref{fig:4c}(b) for $D=6$: (a) $k=0.78427$; (b)
$k=0.78426$. } \label{fig:4e}
\end{figure}

Figure~\ref{fig:4c} shows ES branches in the case of
self-defocusing $\chi ^{(3)}$ terms, with $\gamma _{1}=\gamma
_{2}=-0.05$ and $\delta <0$. Figures~\ref{fig:4d} and \ref{fig:4e}
display the profiles of these ESs for $D=100$, $D=10$ and $D=6$.
Note that these curves and the solutions on them for $D=100 $ are
identical, to the accuracy depicted, to the corresponding ones
found in the continuum counterpart of the model in Ref.
\cite[Figs.~2,3]{YaMaKaCh:01}.
Notice the variety of multi-humped shapes of the solitons
belonging to these families. Like the continuum model, only the
inner-most of these branches represents a fundamental soliton.
Continuation towards the anti-continuum limit, $D\rightarrow 0$,
becomes numerically problematic for these solitons. It was found
difficult to compute the solutions with repeatable accuracy while
varying $N$ below $D\approx 6$. Clearly, these branches become
increasingly spiky as $D$ is decreased (see Fig.~\ref{fig:4e}),
and it may happen that branches of ES solutions actually terminate
before they reach the minimum value of $D$ at which they remain
embedded, which would be $D_{\min}=2k-1$, for the values of $q$
and $\delta $ used in Fig.~\ref{fig:4c}(b).

\begin{figure}[tbp]
\begin{center}
\includegraphics[scale=0.7]{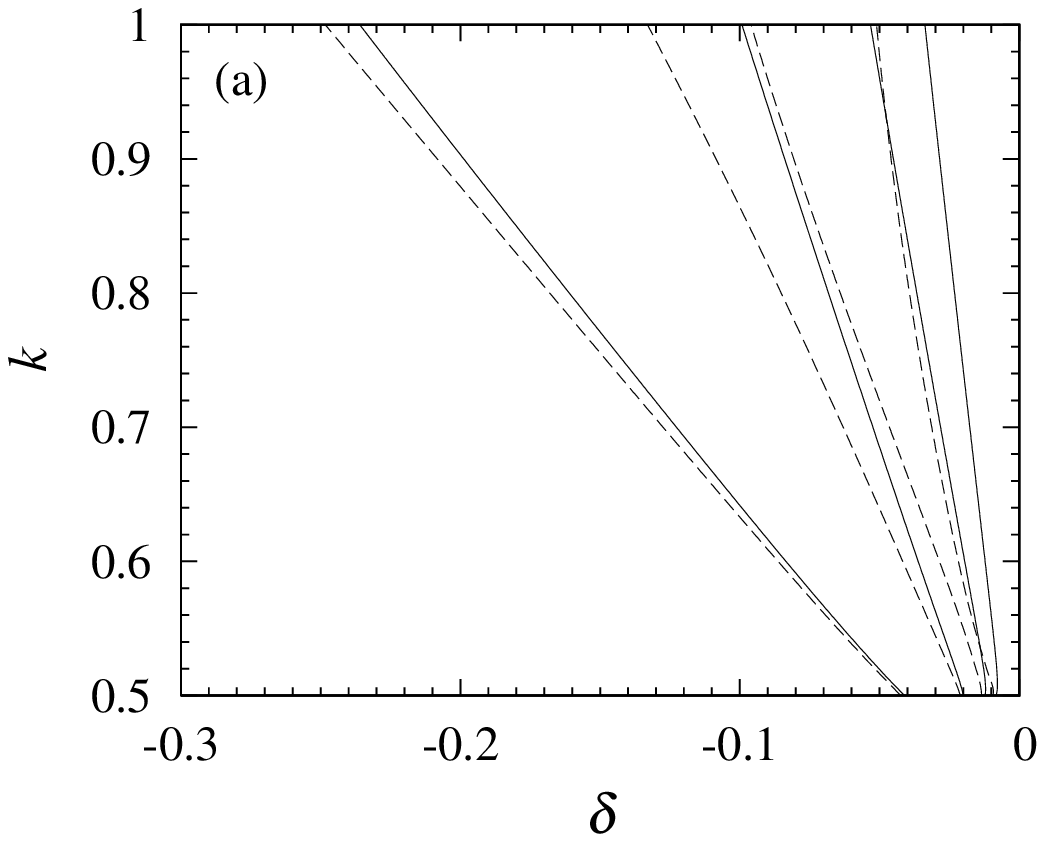}\quad \includegraphics[scale=0.7]{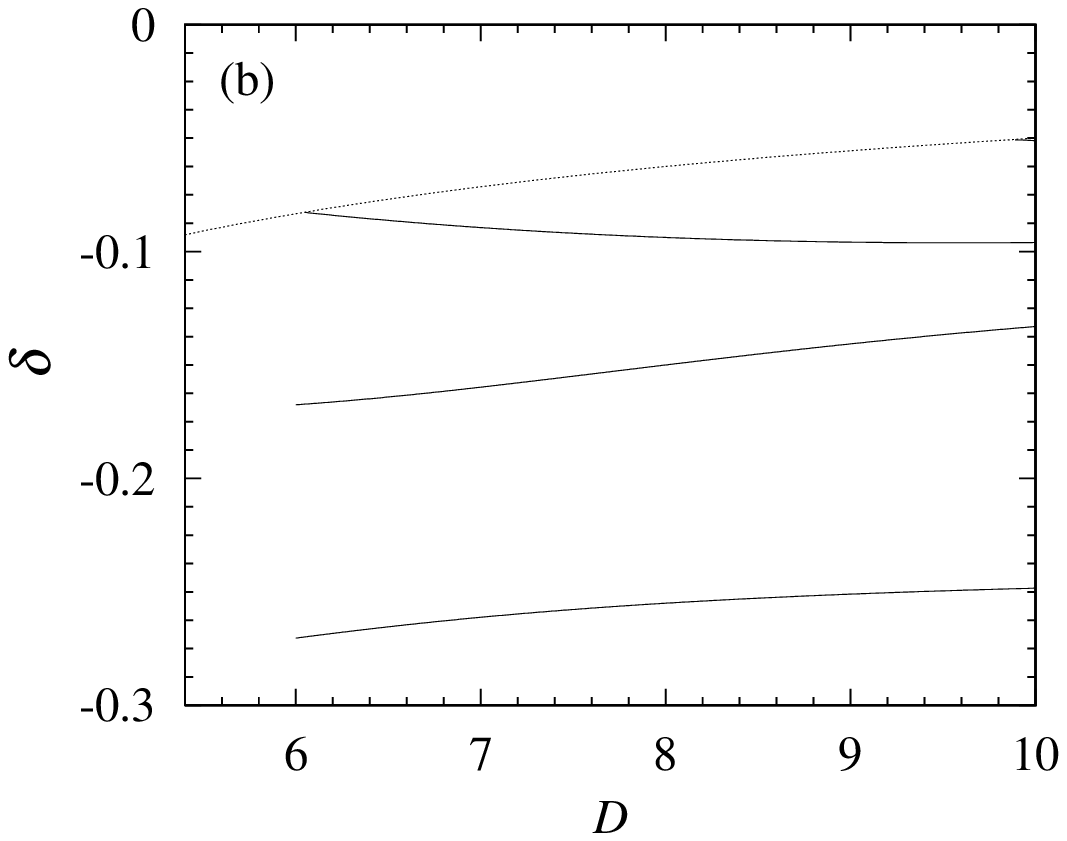}
\end{center}
\caption{Branches of ESs with $q=1$, $\protect\gamma _{1}=0.05$
and $\protect\gamma _{2}=-0.05$: (a) In the $(\protect\delta
,k)$-plane for $D=100$ and $10$; (b) in the $(D,k)$-plane for
$k=1$ and $N=24$. In panel~(a), the solid and dashed curves
represent the results for $D=100$ and $10$, respectively. When
$D=100$ (resp. $D=10$), $N=85$ (resp. $N=30$) was used. In
panel~(b), the forth solid line from the bottom still exists
although it is very short and almost invisible. The ESs exist only
in the region $0.5<k<0.5-\protect\delta D$ when $q=1$, whose
boundary is indicated by a dotted line. } \label{fig:4f}
\end{figure}

\begin{figure}[tbp]
\begin{center}
\includegraphics[scale=0.55]{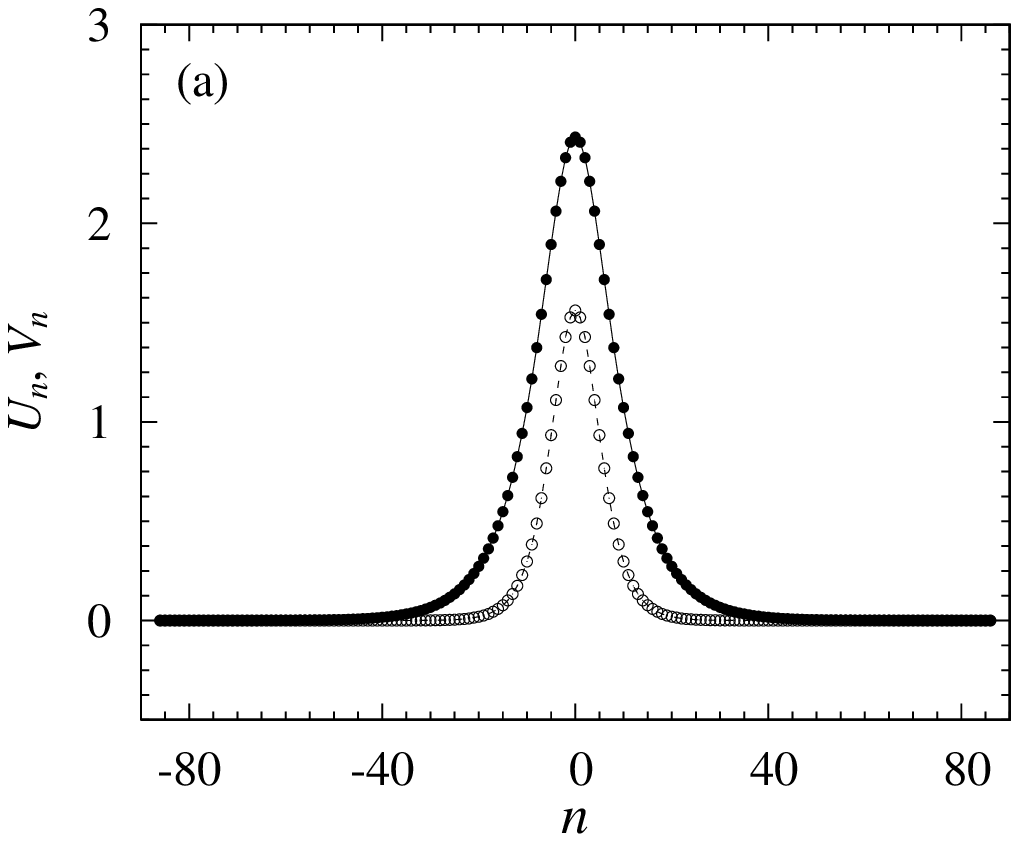}\quad \includegraphics[scale=0.55]{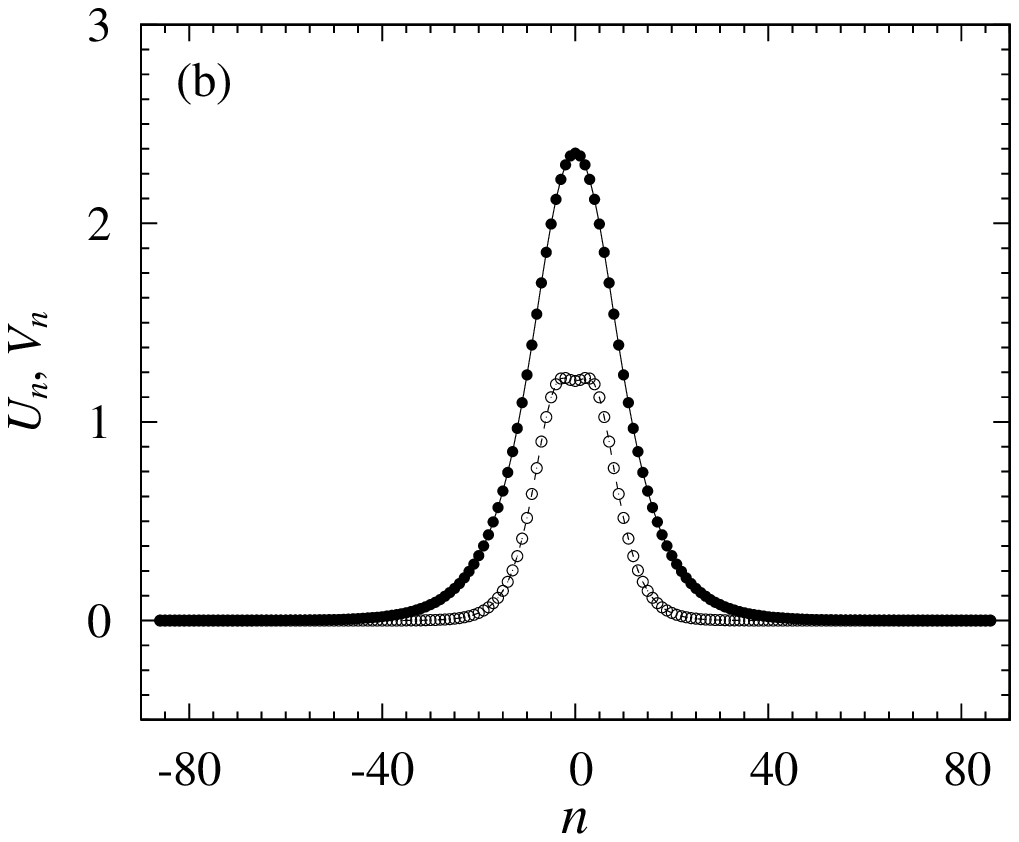}\\[2ex]
\includegraphics[scale=0.55]{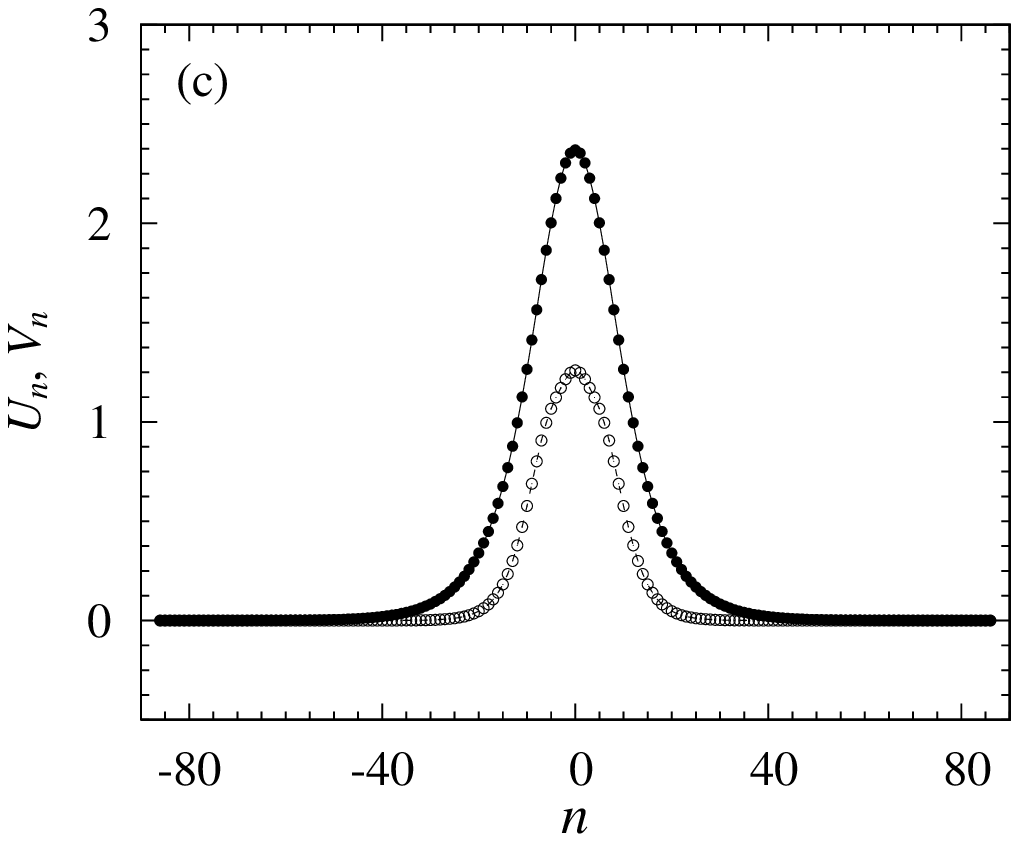}\quad \includegraphics[scale=0.55]{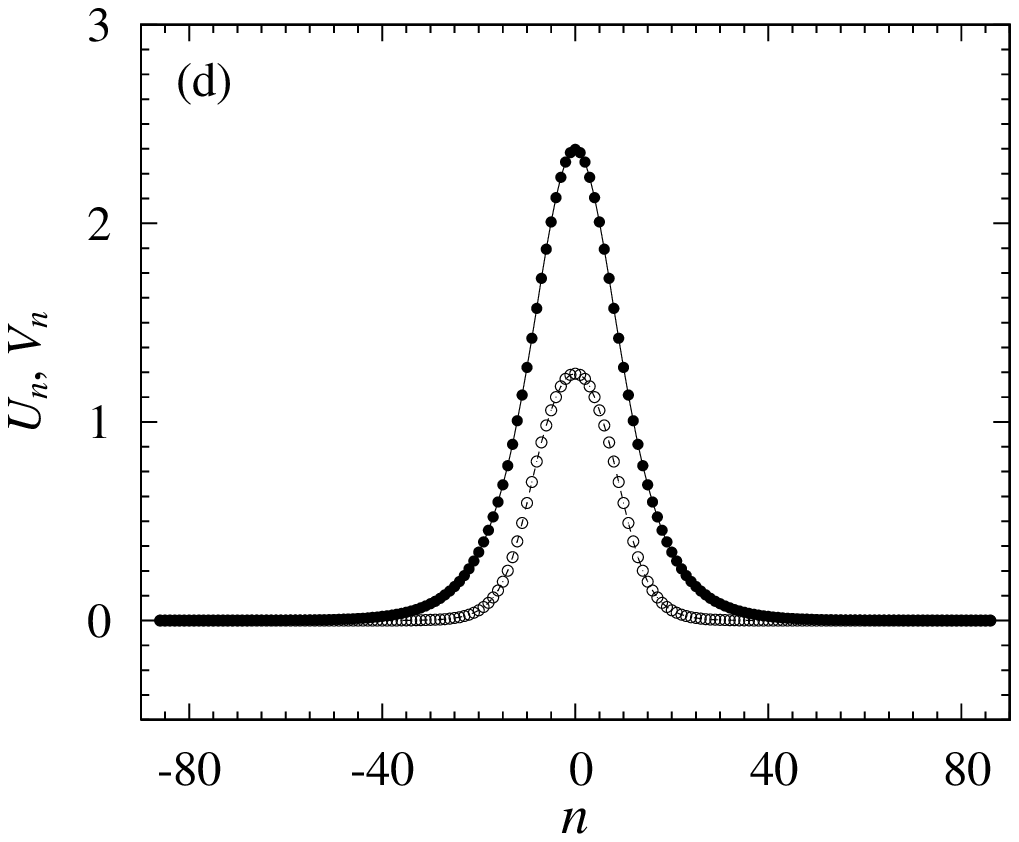}\\[2ex]
\includegraphics[scale=0.55]{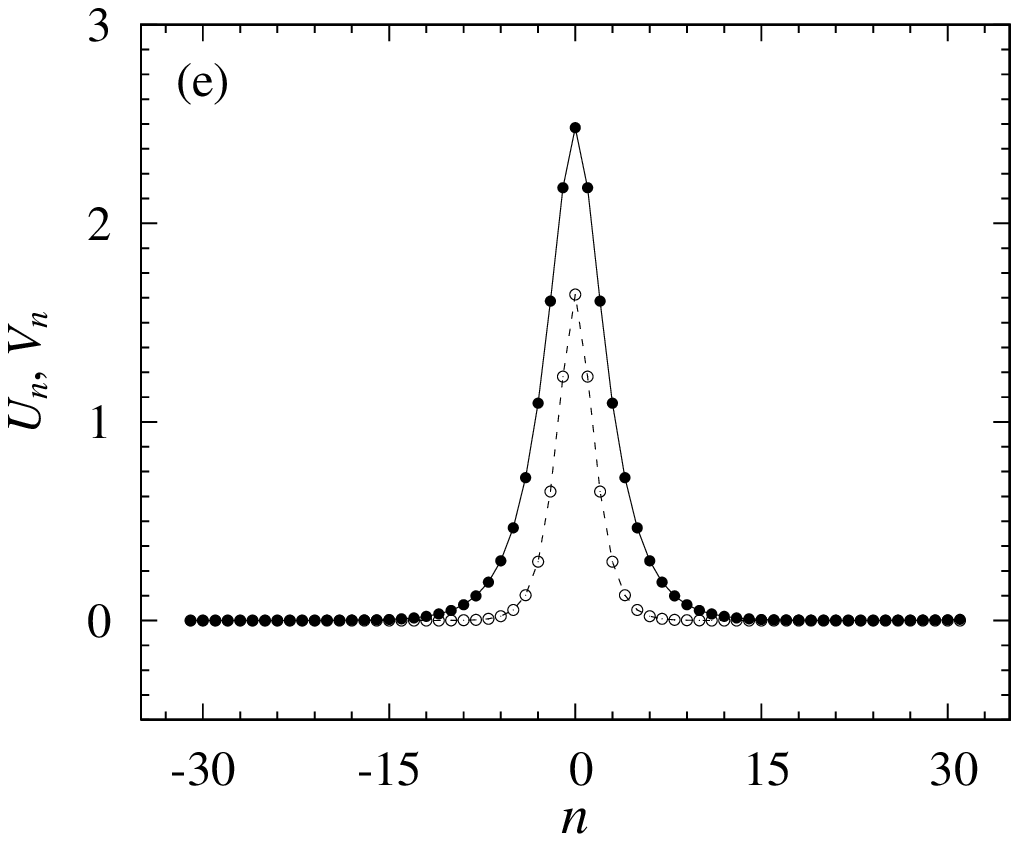}\quad \includegraphics[scale=0.55]{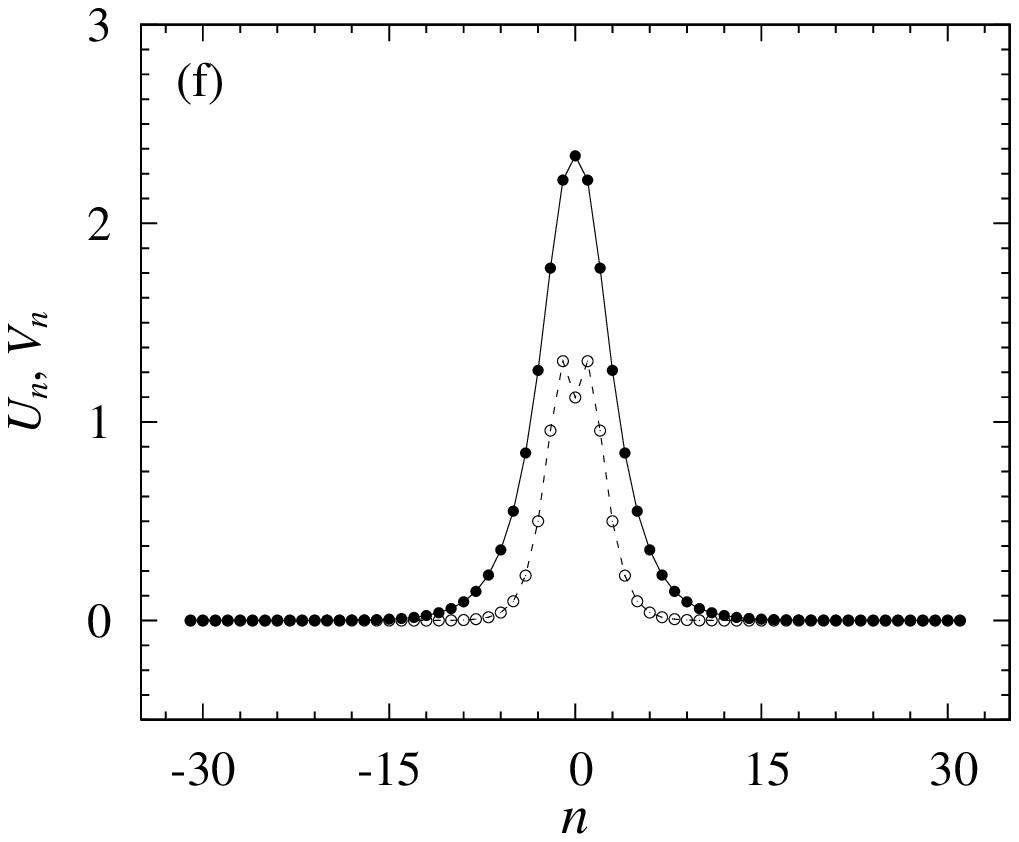}\\[2ex]
\includegraphics[scale=0.55]{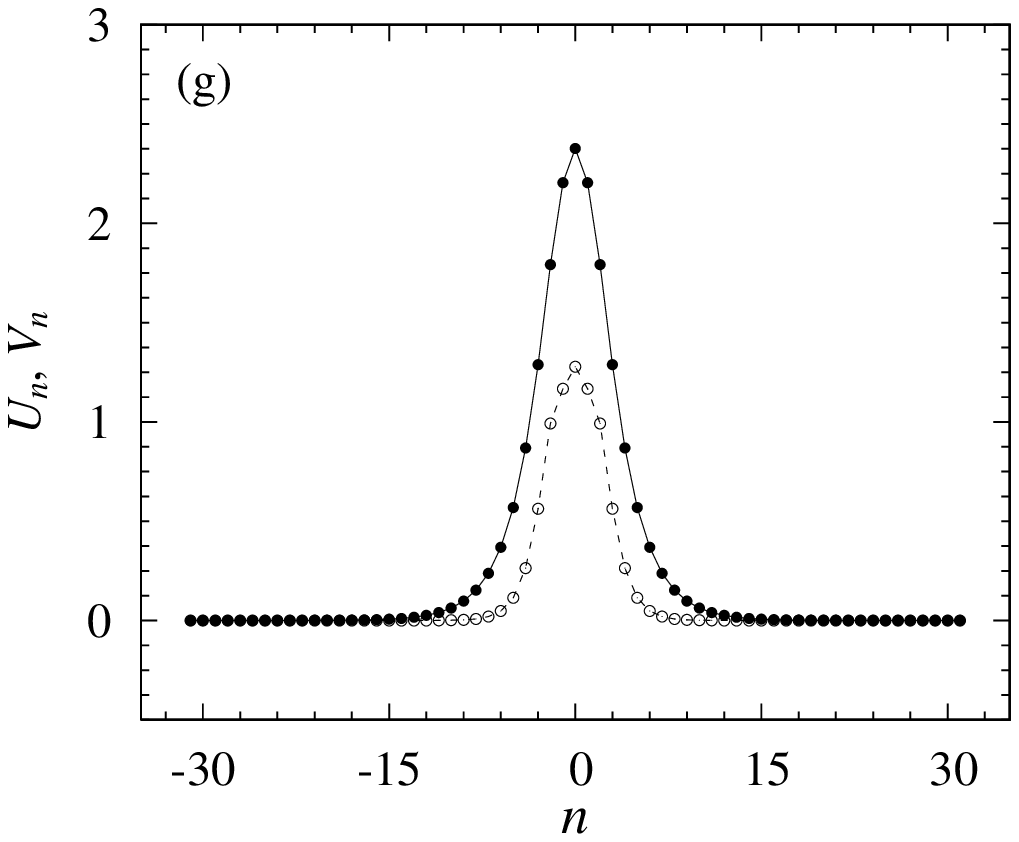}\quad \includegraphics[scale=0.55]{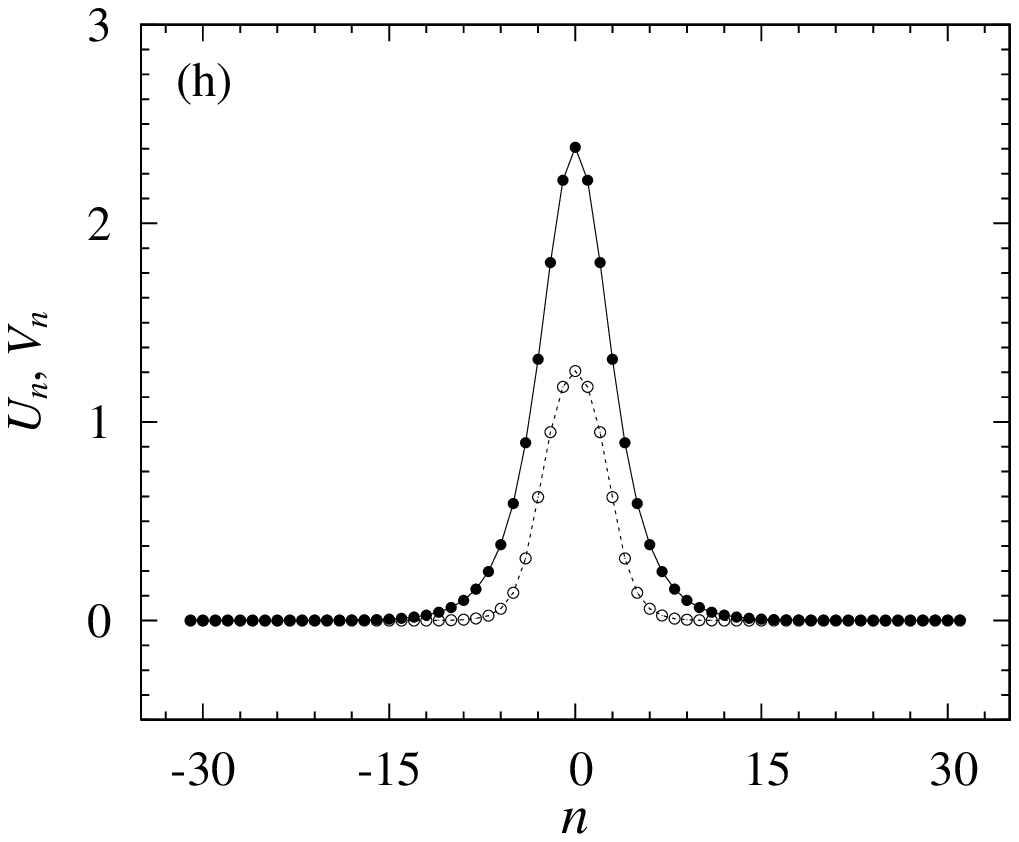}
\end{center}
\caption{Examples of ESs on the branches in
Fig.~\protect\ref{fig:4f}(a) for $k=1$: (a) $\protect\delta
=-0.23601$; (b) $\protect\delta =-0.099245$; (c) $\protect\delta
=-0.053237$; (d) $\protect\delta =-0.033776$; (e) $\protect\delta
=-0.24838$; (f) $\protect\delta =-0.13304$; (g) $\protect\delta
=-0.096003$; (h) $\protect\delta =-0.05109$. In panels (a)-(d),
$D=100$ and $N=85$; in panels (e)-(h), $D=10$ and $N=30$. }
\label{fig:4g}
\end{figure}

Figure~\ref{fig:4f} shows the ES branches in a still more exotic
case of opposite signs in front of the FF and SH $\chi ^{(3)}$ SPM
terms, $\gamma _{1}=0.05$ and $\gamma _{2}=-0.05$. This case may,
in principle, also be physically relevant -- not to ordinary
optical materials, but rather to photonic crystals (see, e.g.,
\cite{PhotCryst} and references therein). Figure~\ref{fig:4g}
displays the profiles of these ESs for $D=100$ and $D=10$. Note
similarity with the branches in Fig.~\ref{fig:4c} for small
$\delta $. However, in this case it is found that the solution
branches still exist for large $k$, rather than undergoing turning
back with the increase of $k$. Also the internal oscillations on
the non-fundamental branches become far less pronounced.

\begin{figure}[t]
\begin{center}
\includegraphics[scale=0.7]{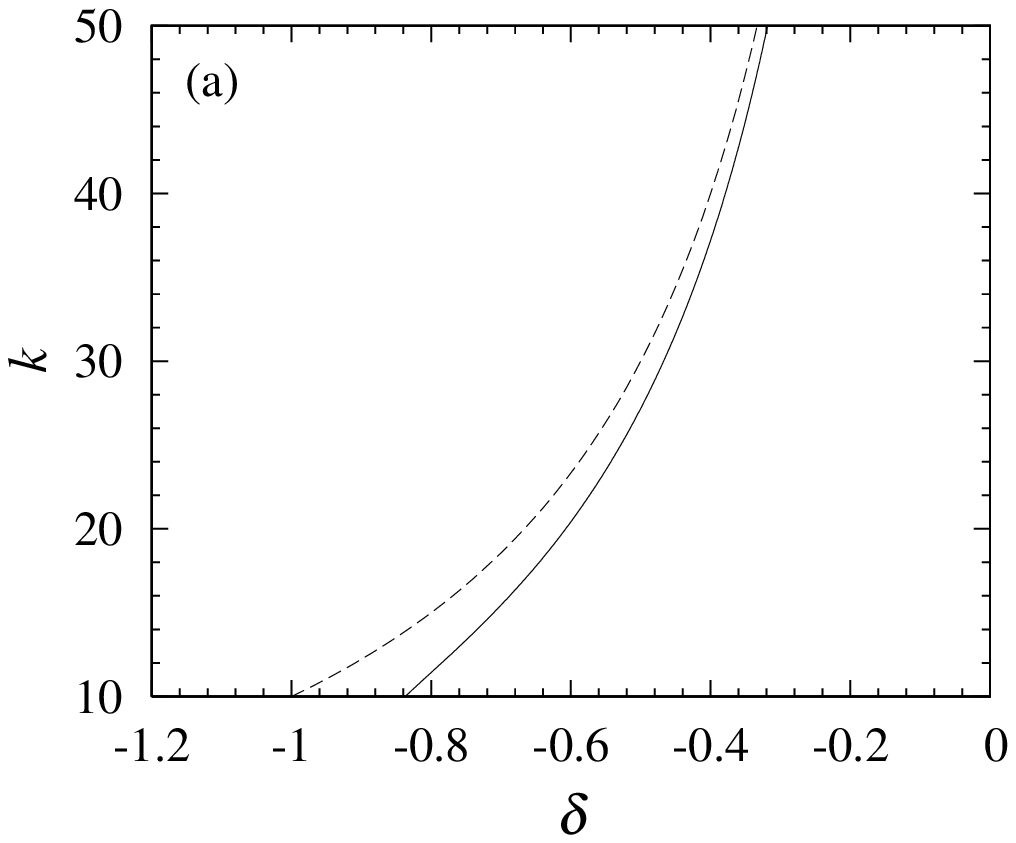}\quad \includegraphics[scale=1.23]{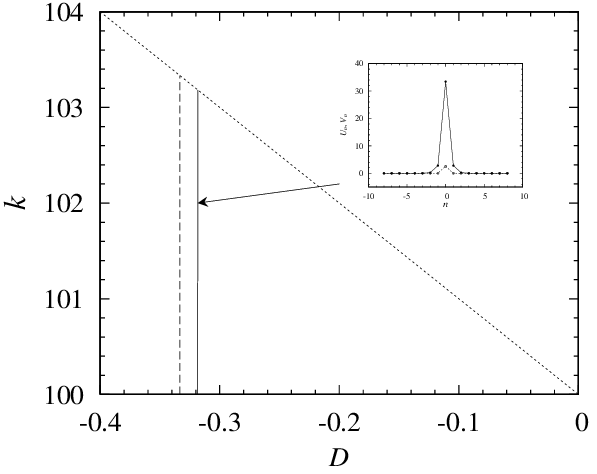}
\end{center}
\caption{Branches of discrete ESs for $\protect\gamma _{1}=0.05$,
$\protect\gamma _{2}=-0.05$, $D=10$ and $N=7$. The analytical
prediction, given by Eqs.~(\protect\ref{final}) and
(\protect\ref{rescaledpars}) is plotted by dashed lines. (a)
Solutions in the $(\protect\delta ,k)$-plane for $q=2k$. (b)
Solutions in the $(\protect\delta ,q)$-plane for $k=50$. The ESs
exist in the region $100+10\protect\delta<q<100$, whose boundary
is shown by the dotted line. An example of the ES, for
$\protect\delta =-0.31873$ and $q=98$, is plotted in panel~(b). }
\label{fig:4h}
\end{figure}

The approximate analysis of Section~3 is also valid for $\gamma
_{1}>0$ and $\gamma _{2},\,\delta <0$ when $k$ is large.
Figure~\ref{fig:4h} shows the fundamental ES branch (the left one
in Fig.~\ref{fig:4f}(a)) with the same values of $\gamma _{1}$ and
$\gamma _{2}$ as those in Fig.~\ref{fig:4f} for $D=10$ when $k$ is
rather large. In Fig.~\ref{fig:4h}(a), the parameters $\delta $
and $k$ are varied for $q=2k$, and in Fig.~\ref{fig:4h}(b) the
parameters $\delta $ and $q$ are varied for $k=50$, as in
Figs.~\ref{fig:4b1}(a) and (b). The predictions based on Eqs.
(\ref{final}) with (\ref{rescaledpars}) are plotted as dashed
lines. A profile of the ES is also displayed in
Fig.~\ref{fig:4h}(a). We see that Eq.~(\ref{final}) again
approximates well the numerical result for the fundamental
solitons when $k$ is large. The ES in Fig.~\ref{fig:4h}(a)
features a steep peak at $n=0$, as assumed in the analytical
approximation.

\begin{figure}[t]
\begin{center}
\includegraphics[scale=1.23]{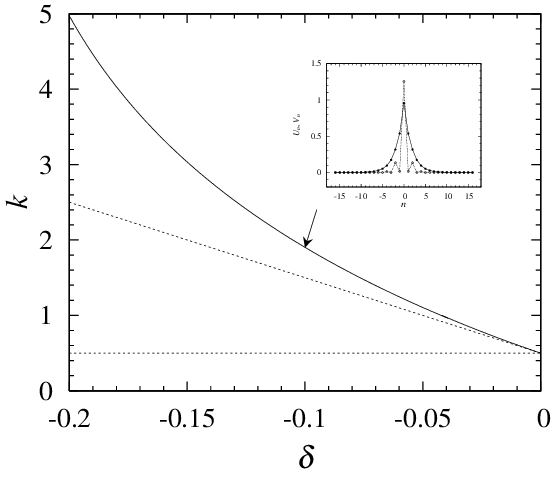}\quad \includegraphics[scale=0.7]{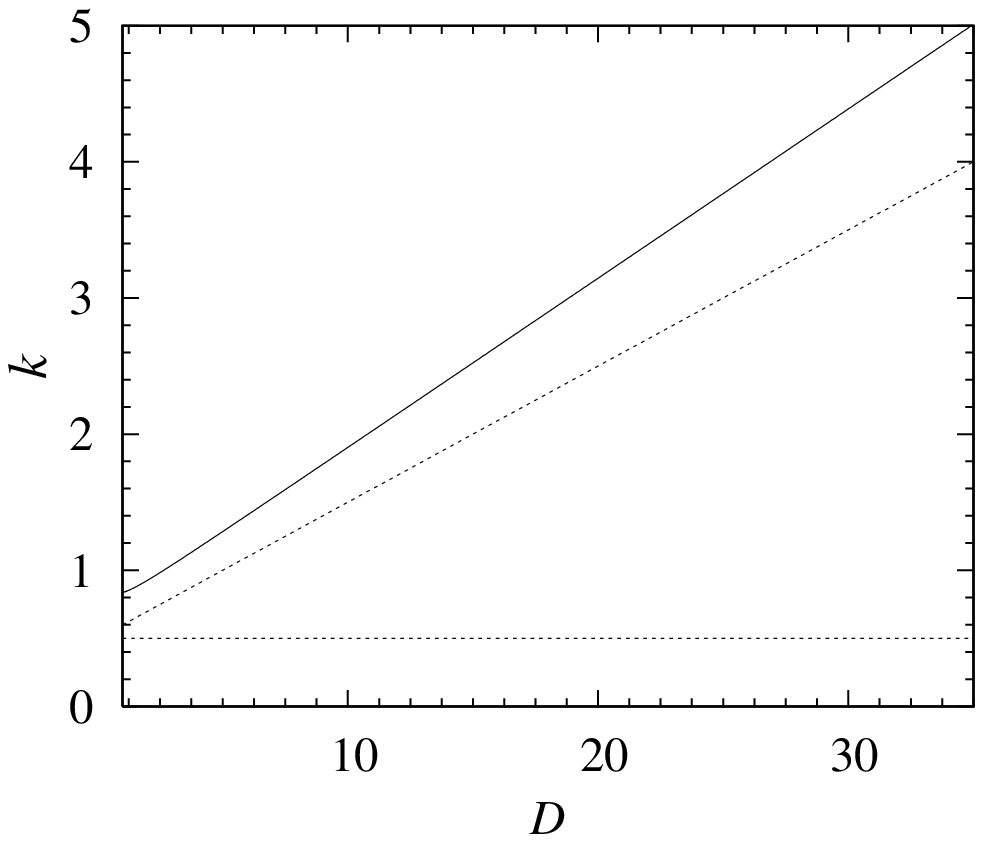}
\end{center}
\caption{Branches of solutions of the three-dimensional algebraic problem
for $G_{\protect\epsilon }$ in the absence of the $\protect\chi ^{(3)}$
nonlinearity, with $\protect\gamma _{1}=\protect\gamma _{2}=0$, and $q=1$.
(a) In the $(\protect\delta ,k)$-plane for $D=10$; (b) in the $(D,k)$-plane
for $\protect\delta =-0.1$ and $N=15$. In panel~(a), $N=15$ and $25$ were
used for $k>1$ and $k<1$, respectively. The ESs would exist only in the
region $0.5<k<0.5-\protect\delta D$, the boundaries of which are plotted as
dotted lines, when $q=1$. An example of a \emph{non-embedded} (gap) soliton,
found in this case for $\protect\delta =-0.1$ and $k=1.9027$, is plotted in
panel~(a).}
\label{fig:4i}
\end{figure}

Finally, we briefly discuss the case in which $\gamma _{1}=\gamma
_{2}=0$, i.e., the $\chi ^{(3)}$ terms are absent. As shown in
Fig.~\ref{fig:4i}, we could follow solutions of the
three-dimensional algebraic problem for $G_{\epsilon}$. However,
these solutions do not represent ESs since the origin is not a
saddle-center but a hyperbolic saddle. Although the region
$q/2<k<q/2-\delta D$ in which the origin is a saddle-center
becomes wider when $D$ is larger, for the solution $k$ diverges to
$\infty$ as $D\rightarrow \infty$. Thus, there seems to be no hope
that an ES exists in this case. As mentioned above, this situation
is very similar to that in the continuum model (\ref{eqn:PDE}).

\section{Conclusions}

In this paper, we have established the existence of embedded
solitons (ESs) in the discrete model of the second-harmonic
generation in the presence of cubic nonlinearity. The model can be
naturally realized as an array of channel waveguides in the
spatial domain, therefore our results suggest possibilities for
new experiments with discrete spatial solitons in nonlinear
optics. We have also introduced a simplified model, with the $\chi
^{(3)}$ nonlinearity present solely at the central site, in which
the existence of an ES was demonstrated in an asymptotic
analytical form. In the general case, the existence region for the
ESs in the discrete model was found numerically. Moreover, we have
checked that the asymptotic analysis of the simplified model in
the limit of large wavenumbers gives a good approximation of the
codimension-one set in parameter space, on which the ESs exist.

More generally, we have established, that unlike the
discretizations of other (dissipative) continuum models bearing
localized solutions, the codimension of ESs does not change when
one passes to a discrete version. Whereas continuum ESs may be
regarded as homoclinic orbits to saddle-center equilibria in
reversible flows (ordinary differential equations), discrete ESs
should be thought of as homoclinic orbits to saddle-center fixed
points of reversible maps. Both have codimension one in the
parameter space. Understanding this property led us to the
derivation of a numerical method for computing homoclinic orbits
to nonhyperbolic equilibria of reversible maps.

Accurate investigation of stability of ESs in the lattice model is
beyond the scope of the present investigation. Systematic results
for the stability will be presented elsewhere; however, some
preliminary results suggest that the fundamental discrete ESs
inherit the \textit{semi-stability} property found in the
corresponding continuous model \cite{YaMaKaCh:01,PeYa:02}. We
expect that general arguments in favor of the semi-stable
character of these solitons for the continuous case should apply
in the discrete setting as well.

\begin{figure}[t]
\begin{center}
\includegraphics[scale=0.8]{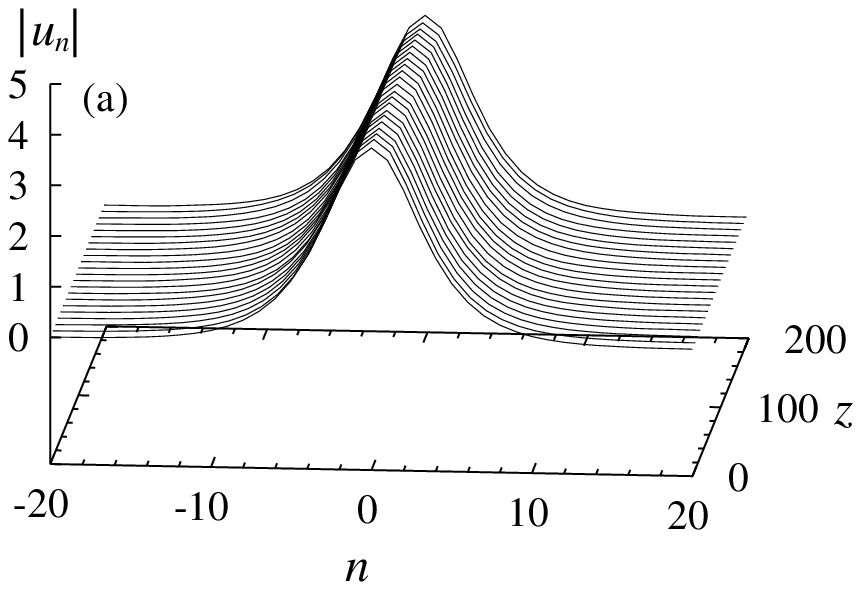}\quad \includegraphics[scale=0.8]{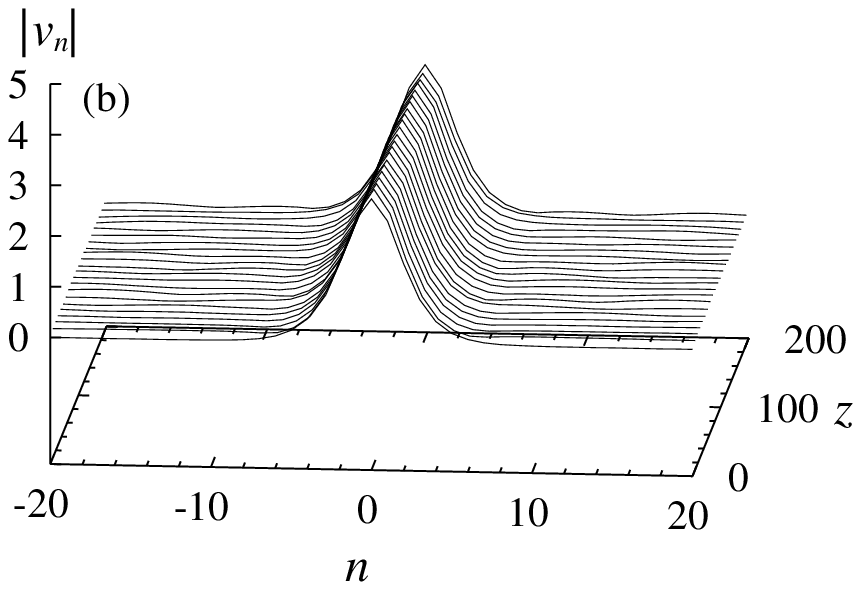}\\[3ex]
\includegraphics[scale=0.6]{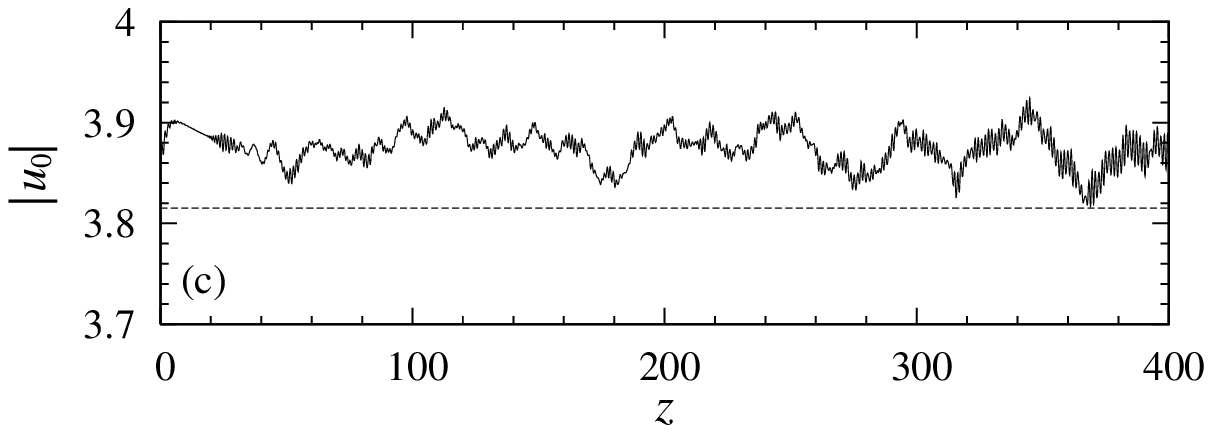}\quad \includegraphics[scale=0.6]{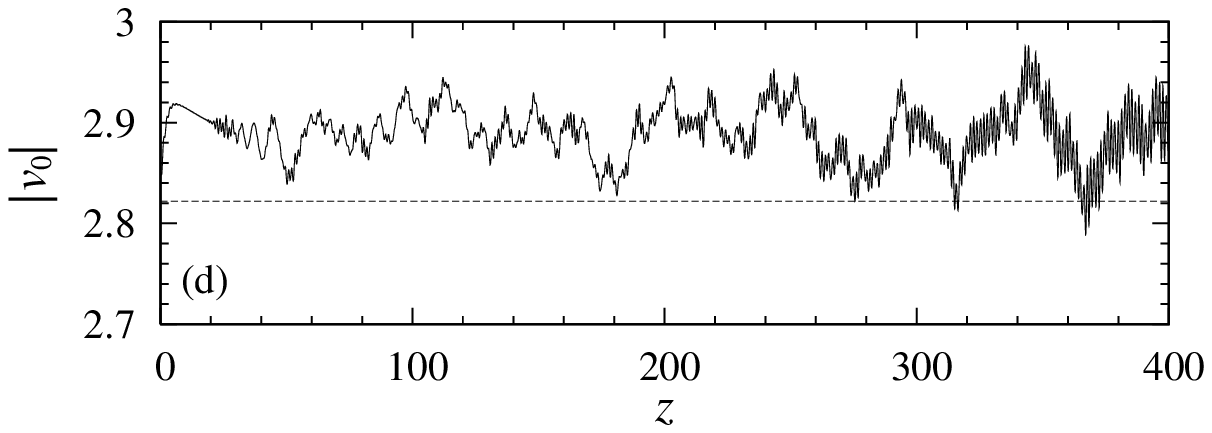}\\[0pt]
\end{center}
\caption{Evolution of the fundamental ES (corresponding to
Fig.~\protect\ref{fig:4d}(e)) initiated by the nondestructive
perturbation with $c_{1}=c_{2}=0.01$ in the initial conditions
(\protect\ref{eqn:init}). The dashed lines in panels~(c) and (d)
are the amplitudes of the $u_{0}$- and $v_{0}$-components of the
unperturbed soliton.} \label{fig:7a}
\end{figure}

Figure~\ref{fig:7a} shows a preliminary computational result for
$D=10$, $\delta =-1$, $k=0.6895$, $q=1$ and $\gamma _{1}=\gamma
_{2}=-0.05$, corresponding to the fundamental ES of
Fig.~\ref{fig:4d}(e). Here Eq.~(\ref{eqn:lattice}) was integrated
numerically by the fourth-order Runge-Kutta method under the
boundary condition
\begin{equation}
u_{-\bar{N}-1}(z)=u_{\bar{N}+1}(z)=v_{-\bar{N}-1}(z)=v_{\bar{N}+1}(z)=0
\end{equation}for all $z\geq 0$ with $\bar{N}=93$ and the initial condition
\begin{equation}
u_{n}(0)=(1+c_{1})U_{n},\quad v_{n}(0)=(1+c_{2})V_{n}  \label{eqn:init}
\end{equation}where $(U_{n},V_{n})$ represents an approximate ES given by the data of
Fig.~\ref{fig:4d}(e) for $|n|\leq N=30$ and by $U_{n}=V_{n}=0$ for $|n|>N$, and $c_{1,2}$
are small amplitudes of the initial disturbance, which were chosen
to be $c_{1}=c_{2}=0.01$. The positive values of $c_{1,2}$ mean that the
norm
\begin{equation}
E=\sum_{n=-\bar{N}}^{\bar{N}}(|u_{n}|^{2}+2|v_{n}|^{2})
\end{equation}of the perturbed state  (in the temporal-domain optical model, it plays the
role of energy) is greater than that of the unperturbed ES. In
Fig.~\ref{fig:7a} we see the characteristic hallmark of
semi-stability for the ES: The shapes of the perturbed wave are
almost unchanged for a long period of $t $ in
Figs.~\ref{fig:7a}(a) and(b), and the amplitudes $|u_{0}(t)|$ and
$|v_{0}(t)|$ exhibit small oscillations near the unperturbed
values in Figs.~\ref{fig:7a}(c) and(d). However, the perturbed ES
was destroyed in the simulations when different signs of $c_{1}$
and $c_{2}$ were chosen. Further investigation of stability and
bifurcation will be the subject of future work.

Finally, our results so far pertain only to those discrete ESs
that are symmetric with respect to the involution $R$ defined in
Eq.~(\ref{R1}); solitons with this symmetry have an on-site
central peak. It is also possible to apply techniques developed in
this work to solutions symmetric with respect to involution
$\hat{R}$, see Eq.~(\ref{R2}), that should feature an inter-site
central peak. In other physical context, such waves are less
likely to be stable than waves that are centered on a lattice site
\cite{Review}. However, that understanding typically applies to
regular (gap) discrete solitons and need not necessarily apply to
ESs. We shall address this issue in future work.

\section*{Acknowledgements}

 K.Y. acknowledges support from the Japanese Society for
the Promotion of Science, which enabled him to stay in Bristol and
to perform this work. B.A.M. appreciates support of EPSRC
``critical mass" grant and mathematics programme and hospitality
of the Department of Engineering Mathematics at the University of
Bristol.

\section*{References}


\end{document}